\pdfoutput=1 

\documentclass[acmsmall,screen]{acmart}

\setcopyright{none}
\acmConference{}
\acmBooktitle{}
\acmPrice{}
\acmISBN{}

\renewcommand\footnotetextcopyrightpermission[1]{} %

\usepackage[utf8]{inputenc}
\usepackage[english]{babel}
\usepackage{xcolor}
\usepackage{xspace}
\usepackage[ruled, vlined, linesnumbered, commentsnumbered, longend]{algorithm2e}

\SetCommentSty{mycommfont}
\usepackage{subfigure}
\usepackage{enumitem}
\usepackage{longfbox}
\usepackage{tabularx}
\usepackage{multirow}
\usepackage{adjustbox}
\usepackage{array}
\usepackage{graphicx}
\usepackage{graphics}
\usepackage[normalem]{ulem}
\usepackage{tabulary}
\newcolumntype{K}[1]{>{\centering\arraybackslash}p{#1}} %

\usepackage{tikz}

\usetikzlibrary{calc,fit}

\makeatletter
\tikzset{%
  remember picture with id/.style={%
    remember picture,
    overlay,
    save picture id=#1,
  },
  save picture id/.code={%
    \edef\pgf@temp{#1}%
    \immediate\write\pgfutil@auxout{%
      \noexpand\savepointas{\pgf@temp}{\pgfpictureid}}%
  },
  if picture id/.code args={#1#2#3}{%
    \@ifundefined{save@pt@#1}{%
      \pgfkeysalso{#3}%
    }{
      \pgfkeysalso{#2}%
    }
  }
}

\def\savepointas#1#2{%
  \expandafter\gdef\csname save@pt@#1\endcsname{#2}%
}

\def\tmk@labeldef#1,#2\@nil{%
  \def\tmk@label{#1}%
  \def\tmk@def{#2}%
}

\tikzdeclarecoordinatesystem{pic}{%
  \pgfutil@in@,{#1}%
  \ifpgfutil@in@%
    \tmk@labeldef#1\@nil
  \else
    \tmk@labeldef#1,(0pt,0pt)\@nil
  \fi
  \@ifundefined{save@pt@\tmk@label}{%
    \tikz@scan@one@point\pgfutil@firstofone\tmk@def
  }{%
  \pgfsys@getposition{\csname save@pt@\tmk@label\endcsname}\save@orig@pic%
  \pgfsys@getposition{\pgfpictureid}\save@this@pic%
  \pgf@process{\pgfpointorigin\save@this@pic}%
  \pgf@xa=\pgf@x
  \pgf@ya=\pgf@y
  \pgf@process{\pgfpointorigin\save@orig@pic}%
  \advance\pgf@x by -\pgf@xa
  \advance\pgf@y by -\pgf@ya
  }%
}
\newcommand\tikzmark[2][]{%
\tikz[remember picture with id=#2] #1;}
\makeatother

\newcommand\MyBox[4][-1ex]{%
  \tikz[remember picture,overlay,pin distance=0cm]
  {\draw[draw=#4,line width=1pt,fill=#4!20,rectangle,rounded corners]
( $ (pic cs:#2) + (-1ex,2ex) $ ) rectangle ( $ (pic cs:#3) + (1ex,#1) $ );
}
}
\newcommand\MyBoxH[4][-1ex]{%
  \tikz[remember picture,overlay,pin distance=0cm]
  {\draw[draw=#4,line width=1pt,fill=#4!20,rectangle,rounded corners]
( $ (pic cs:#2) + (-4.5ex,2ex) $ ) rectangle ( $ (pic cs:#3) + (1ex,#1) $ );
}
}

\usepackage{listings}
\usepackage[most]{tcolorbox}

\newcommand{\RevDel}[1]{}
\newcommand{\RevAdd}[2]{#2\xspace}

\newcommand{\Comment}[1]{}
\newcommand{\parabf}[1]{\noindent\textbf{#1}}
\newcommand{\type}[1]{\textbf{\texttt{#1}}}
\newcommand{\cons}[1]{\texttt{#1}}
\newcommand{\Constructor}{\emph{constructor}\xspace}

\newcommand{\NBugs}{\RevDel{40}\RevAdd{}{49}\xspace}     %
\newcommand{\NFix}{\RevDel{24}\RevAdd{}{25}\xspace}      %
\newcommand{\NConfirm}{\RevDel{30}\RevAdd{}{37}\xspace}  %
\newcommand{\NIRBug}{17\xspace}    %
\newcommand{\MoreBug}{\RevDel{6.7x}\RevAdd{}{6.16x}\xspace}

\newcommand{\pspec}{domain-specific\xspace} %
\newcommand{\PSpec}{Domain-Specific\xspace} 
\newcommand{\NPSpecMut}{3\xspace} %

\newcommand{\RunTVM}{\textsc{ExecuteTVM}}

\newcommand{\cppLine}{$\sim$150\xspace}

\usepackage{pifont}
\newcommand{\Y}{\ding{51}}%
\newcommand{\blackding}[1]{\ding{\numexpr181+#1\relax}}

\definecolor{Del}{rgb}{0.99, 0.85, 0.85}
\definecolor{HLRed}{rgb}{0.95, 0.47, 0.55} %
\definecolor{Add}{rgb}{0.77, 0.93, 0.79}
\definecolor{HLGreen}{rgb}{0.35, 0.85, 0.35}

\newcommand{\AddLine}[1]{\noindent\colorbox{Add}{\makebox[\dimexpr\linewidth-2\fboxsep][l]{\textcolor{HLGreen}{+} #1}}}
\newcommand{\DelLine}[1]{\noindent\colorbox{Del}{\makebox[\dimexpr\linewidth-2\fboxsep][l]{\textcolor{HLRed}{-} #1}}}

\newfboxstyle{hlred}{padding=1pt,margin=0pt,baseline-skip=false,background-color=HLRed,border-color=HLRed}
\newfboxstyle{hlgreen}{padding=1pt,margin=0pt,baseline-skip=false,background-color=HLGreen,border-color=HLGreen}

\newcommand{\File}{F}
\newcommand{\PassSeq}{P}
\newcommand{\NFail}{N}
\newcommand{\Time}{T}
\newcommand{\Pool}{\mathcal{S}}
\newcommand{\Null}{NULL\xspace}

\newcommand\tstatesym S
\newcommand\tprimsym R
\newcommand\tsym T
\newcommand\iprimsym P
\newcommand\isym I

\newcommand\Mut M

\newcommand{\irTypeCorpus}{\begin{equation}
    \mathbf{NodeTypes} = \{\mathit{NodeType}_1, \mathit{NodeType}_2, \dots, \mathit{NodeType}_n\}.
\end{equation}}

\newcommand{\irConstructorDefinition}{\begin{equation}
    (\mathit{NodeType}_{i_1}, \mathit{NodeType}_{i_2}, \dots, \mathit{NodeType}_{i_p}) \to \mathit{NodeType}_{i_r}.
\end{equation}}

\newcommand{\irExampleDefinition}{\begin{equation}
    \cons{PrimFunc}([\mathtt a, \mathtt b], \cons{While}(\cons{And}(\cons{EQ}(\cons{a}, \cons{5}), \cons{GT}(\cons{b}, \cons{3})), \cons{...})): \type{PrimFunc}.
\end{equation}}

\newcommand{\irExampleContext}{\begin{equation}
    \cons{PrimFunc}([\cons a, \cons b], \cons{While}(\square{}, \cons{...})): \mathit{Context}.\label{eq:irExampleContext}
\end{equation}}

\newcommand\irExampleConstraints{\begin{equation}
    (\type{PrimExpr}, \cons{bool}, [\cons{a}, \cons{b}], [], \mathtt{true}): \mathit{Constraints}.
\end{equation}}

\newcommand\irMutationDefinition{\begin{equation}
    (\mathit{AnyNodeType}, \mathit{Constraints}) \to \mathit{AnyNodeType},
\end{equation}
where $\mathit{AnyNodeType}$ is a disjoint union of all possible $\mathit{NodeType} \in \mathbf{NodeTypes}$, i.e., 
\begin{equation}
    \mathit{AnyNodeType} = \bigsqcup{\mathbf{NodeTypes}}.
\end{equation}}

\newcommand{\sys}{\textsc{Tzer}\xspace}
\newcommand{\tvmfuzz}{\textsc{TVMFuzz}\xspace}
\newcommand{\lemon}{\textsc{LEMON}\xspace}
\newcommand{\deepbillboard}{DeepBillboard\xspace}
\newcommand{\deeptest}{DeepTest\xspace}
\newcommand{\deeproad}{DeepRoad\xspace}
\newcommand{\deepxplore}{DeepXplore\xspace}

\begin{document}
\title{Coverage-Guided Tensor Compiler Fuzzing with Joint IR-Pass Mutation}
\author{Jiawei Liu}
    \affiliation{\institution{University of Illinois at Urbana-Champaign}\country{USA}}
    \email{jiawei6@illinois.edu}
\author{Yuxiang Wei}
    \affiliation{\institution{Tongji University}\country{China}}
    \email{nolest@tongji.edu.cn}
\author{Sen Yang}
    \affiliation{\institution{Fudan University}\country{China}}
    \email{syang15@fudan.edu.cn}
\author{Yinlin Deng}
    \affiliation{\institution{University of Illinois at Urbana-Champaign}\country{USA}}
    \email{yinlind2@illinois.edu}
\author{Lingming Zhang}
    \affiliation{\institution{University of Illinois at Urbana-Champaign}\country{USA}}
    \email{lingming@illinois.edu}
\begin{abstract} %

In the past decade, Deep Learning (DL) systems have been widely deployed in various application domains to facilitate our daily life, e.g., natural language processing, healthcare, activity recognition, and autonomous driving. Meanwhile, it is extremely challenging to ensure the correctness of DL systems (e.g., due to their intrinsic nondeterminism), and bugs in DL systems can cause serious consequences and may even threaten human lives. 
In the literature, researchers have explored various techniques to test, analyze, and verify DL models, since their quality directly affects the corresponding system behaviors.  
Recently, researchers have also proposed novel techniques for testing the underlying operator-level DL libraries (such as TensorFlow and PyTorch), which provide general binary implementations for each high-level DL operator and are the foundation for running DL models on different hardware platforms. 
However, there is still limited work targeting the reliability of the emerging tensor compilers (also known as DL compilers), which aim to automatically compile high-level tensor computation graphs directly into high-performance binaries for better efficiency, portability, and scalability than traditional operator-level libraries. 
Therefore, in this paper, we target the important problem of tensor compiler testing, and have proposed \sys, a practical fuzzing technique for the widely used TVM tensor compiler. \sys focuses on mutating the low-level Intermediate Representation (IR) for TVM due to the limited mutation space for the high-level IR. More specifically, \sys leverages both general-purpose and tensor-compiler-specific mutators guided by coverage feedback for diverse and evolutionary IR mutation; furthermore, since tensor compilers provide various \emph{passes} (i.e., transformations) for IR optimization, \sys also performs pass mutation in tandem with IR mutation for more effective fuzzing.
Our experimental results show that \sys substantially outperforms existing fuzzing techniques on tensor compiler testing, with 75\% higher coverage and 50\% more valuable tests than the 2nd-best technique.
Also, different components of \sys have been validated via ablation study. To date, \sys has detected \NBugs previously unknown bugs for TVM, with \NConfirm bugs confirmed and \NFix bugs fixed (PR merged). %
\end{abstract}
\keywords{Fuzzing, Compiler Testing, Machine Learning Systems}
\maketitle

\section{Introduction}

With the recent advance of deep learning (DL), DL systems have been pervasively deployed in various application domains to facilitate our daily life, including natural language processing~\cite{nlp,bert,vaswani2017attention}, healthcare~\cite{miotto2018deep,esteva2019guide}, activity recognition~\cite{cao2019openpose,kreiss2019pifpaf,guo2021fast}, and autonomous driving~\cite{rao2018deep,grigorescu2020survey}.
Meanwhile, it is extremely challenging to ensure the correctness of DL systems (e.g., due to their intrinsic nondeterminism), and any bug in such decision-making systems can potentially bring serious consequences or accidents (e.g., the life-threatening autonomous-driving failures~\cite{autodrivecomprehensive}).

To date, a large body of prior work has been dedicated to testing, analyzing, and verifying DL models since their quality directly affects the behaviors of DL systems. For example, various techniques have been designed to generate adversarial or edge-case model inputs for testing DL models, including \deepxplore~\cite{deepxplore}, \deeptest~\cite{deeptest},
\deeproad~\cite{deeproad}, TensorFuzz~\cite{odena2019tensorfuzz}, and \deepbillboard~\cite{deepbillboard}. In recent years,
in addition to the algorithmic/model aspect,
researchers also realized the importance of ensuring the correctness of the underlying DL infrastructure supports, and have proposed novel techniques~\cite{lemon, cradle} specifically targeting operator-level DL libraries, such as TensorFlow~\cite{tensorflow} and PyTorch~\cite{pytorch}.
Meanwhile, computation-intensive DL models are being developed everywhere nowadays; early operator-level libraries, which usually only provide a fixed binary for a limited number of platforms, are hardly generalizable and scalable.
Therefore, DL engineers and researchers have been building an ultimate solution, tensor compilers~\cite{tvm, mlir, halide, glow} (also known as DL compilers), to essentially tackle the challenges in performance, portability, and flexibility.
However, to our best knowledge, there is limited work specifically targeting the reliability of the emerging tensor compilers.%

Ensuring the correctness and reliability of tensor compilers is essential for the rise of compilation-based DL infrastructure.
Nonetheless, the complicated software stack of tensor compilers makes it non-trivial for writing hand-crafted unit tests.
For example, in TVM~\cite{tvm} (one of the biggest and most widely used tensor compiler projects), there are over 117k lines of Python code specifically targeting unit testing!%
Designing automated testing techniques for tensor compilers is important but also quite challenging. First, the compiler stack is \emph{deep}, meaning that an input model needs to be compiled through various phases (including numerous parsing, lowering, and optimization passes) to produce the final target code.
Second, the compiler stack is \emph{wide}, meaning that there are innumerable possibilities for composing a single intermediate representation (IR) file or an optimization sequence, let alone their combinations if taking various targets and execution backends into account.

Although some existing fuzzing techniques can potentially be adopted for testing tensor compilers, they are not able to handle the complex compiler infrastructure well. For example, general-purpose binary fuzzers~\cite{afl, libfuzzer} can hardly generate syntactically- and semantically-valid inputs, wasting the majority of time fuzzing the lexical parsing components.
Prior operator-level DL-library testing techniques~\cite{lemon} systematically mutate on the input model seeds to generate diverse model architectures, and can potentially be generalized to most DL infrastructures; however, they are not tailored for tensor compiler testing as they do not consider triggering different optimizations and are also too coarse-grained to generate light-weight yet valuable inputs (as demonstrated by our experimental results in \S~\ref{sec:exp:rq1}).
To our best knowledge, the only existing work specifically targeting tensor compiler fuzzing, \tvmfuzz ~\cite{tvmfuzz}, employs a generation-based approach%
to automatically generate arbitrary low-level IRs for fuzzing TVM.%
However, it suffers from the common limitations of generation-based fuzzing techniques~\cite{holler2012fuzzing, yang2011finding}, e.g., it is challenging to simulate realistic programs to cover deep code paths and the fuzzing process lacks valid guidance; also, it fails to consider the rich search space of possible optimization pass sequences for tensor compilers. As a result, it could only find out very shallow front-end bugs and its coverage growth converges at an early stage (as also confirmed by our experimental results).

In this paper, we focus on practical tensor compiler fuzzing and have made the following design choices. First, we target low-level IR mutation due to the coarse-grained and limited mutation space for high-level IR mutation~\cite{lemon}. Second, we propose the first coverage-guided fuzzing approach for testing tensor compilers, as coverage feedback has been demonstrated to be powerful for exploring deep code paths efficiently in general~\cite{li2018fuzzing}. Following traditional coverage-guided fuzzers~\cite{afl, libfuzzer}, in each iteration, we randomly choose an IR file from a seed pool for mutation and add the newly mutated IR file into the pool only when it triggers new coverage. Meanwhile, instead of relying on the bit-level mutators widely adopted in traditional fuzzers, we develop a set of general-purpose and tensor-compiler-specific mutators for more targeted and effective IR mutation. Third, since a large number of optimization passes can form a pass sequence and potentially be applied to the same IR file to trigger different compiler behaviors, we further build a novel coverage-guided fuzzing strategy to perform joint mutations of both IR and optimization passes for more exhaustive tensor compiler testing. Although our design is general for different tensor compilers, in this paper, we mainly focus on the TVM compiler and have implemented a practical TVM fuzzing technique named \sys. To evaluate the effectiveness of \sys, we have performed an extensive study to compare \sys against LibFuzzer~\cite{libfuzzer} (a state-of-the-art general-purpose fuzzer), \lemon~\cite{lemon} (a state-of-the-art high-level IR fuzzer for DL libraries), and \tvmfuzz~\cite{tvmfuzz} (the only existing low-level IR fuzzer for TVM). Furthermore, we have rigorously evaluated the importance and necessity for all the design choices of \sys.
In summary, the primary contributions of this work go as follows:
\begin{itemize}
    \item \textbf{Novelty}: This paper presents the first coverage-guided fuzzing technique specifically targeting tensor compilers. More specifically, we have designed various general-purpose and tensor-compiler-specific mutators as well as the joint mutation of both IR and optimization passes for effective tensor compiler fuzzing.
    \item \textbf{Implementation}: We have implemented the proposed technique as a practical fuzzer (named \sys) for the TVM compiler. \sys is mainly implemented by over 8.7k lines of Python code together with \cppLine lines of C++ code for extending the LLVM Coverage Sanitizer.
    \sys has been open-sourced at: \url{https://github.com/Tzer-AnonBot/tzer}.
    \item \textbf{Study}: We have performed an extensive study to compare \sys against existing fuzzers for testing TVM, and have also rigorously validated the contribution of each component of \sys. The experimental results show that \sys is able to substantially outperform state-of-the-art fuzzers with 75\% higher coverage and 50\% more valuable tests compared with the 2nd-best fuzzer. Furthermore, different components of \sys all contribute to its final effectiveness. To date, among \NBugs unique new bugs\footnote{We count the number of bugs by unique root fixes (see \S~\ref{sec:metric}).} found by \sys, \NConfirm bugs have been confirmed and \NFix of them have been fixed and merged to the main branch of TVM.
\end{itemize}

\section{Background and Related Work}

\subsection{Tensor Compilers}

\begin{figure}[ht]
    \centering
    \includegraphics[width=\textwidth]{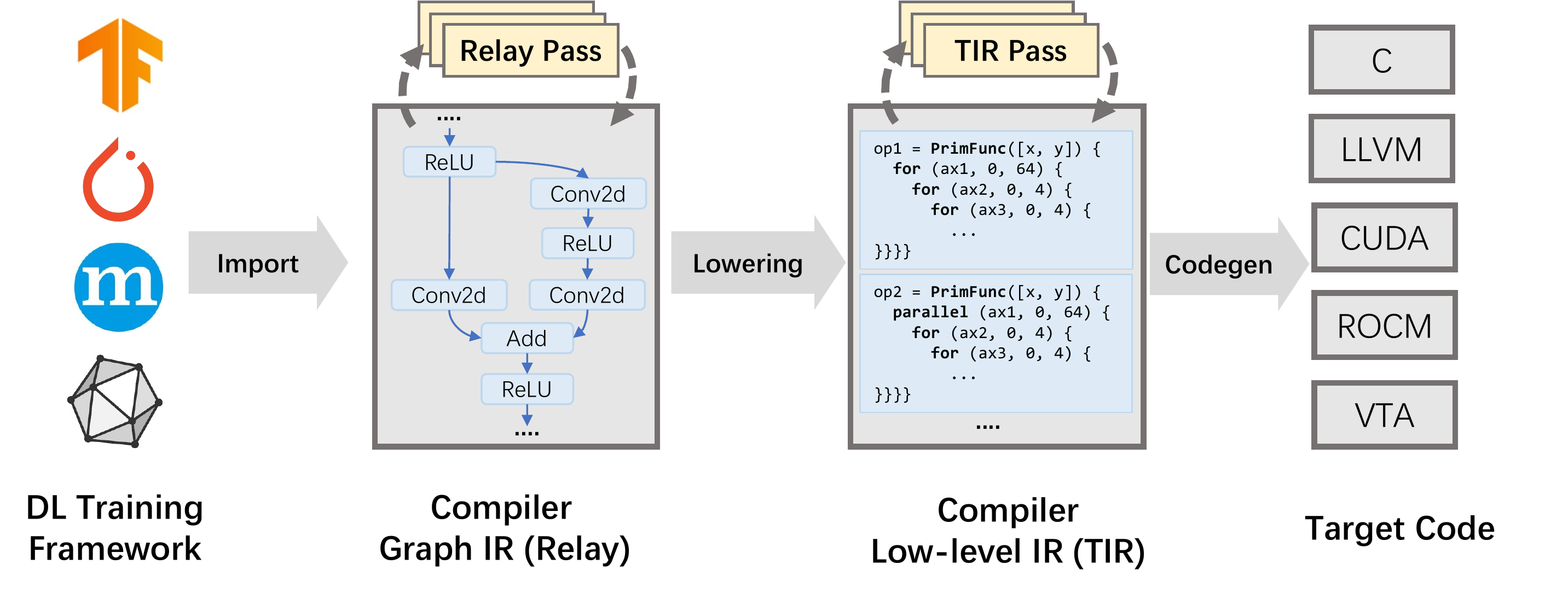}
    \caption{Compilation Flow of TVM.}
    \label{fig:tensor_compiler}
\end{figure}

The computation of deep learning models can be logically described in the dataflow model~\cite{wongsuphasawat2017visualizing}, which is commonly called the \emph{computation graph}~\cite{jia2019taso}.
A computation graph consists of a number of operators (e.g., convolution, max pooling, and many other tensor operations), each of which transforms one or multiple input tensors (i.e., multi-dimensional arrays) into a series of output tensors.
Given the computation graph description, 
there are mainly two approaches for existing DL software to compute it.
Previously, for fast software delivery, ML engineers implemented various operator-level DL libraries, such as TensorFlow~\cite{tensorflow} and PyTorch~\cite{pytorch}, whose operators are implemented with fixed and hand-optimized kernel functions.
However, hand-crafted optimization is time-consuming in the long run and a fixed binary cannot meet the ultimate performance requirements for all hardware vendors.
Therefore, to fundamentally resolve those challenges, recently DL infrastructures have been focusing on developing tensor compilers ~\cite{tvm,xla,glow,plaidml,jin2020compiling,zhao2021akg,tillet2019triton} to automatically generate best-in-class target code for different vendors or even architectures.

Figure \ref{fig:tensor_compiler} illustrates the compilation flow of TVM~\cite{tvm}, one of the most widely-used and advanced tensor compilers (other tensor compilers including XLA~\cite{xla} and Glow~\cite{glow} also follow such logical flow).
First, tensor compilers will transform 3rd-party model files into their own graph representation (i.e., Relay IR in TVM).
Furthermore, a sequence of optimizations (known as \emph{passes} or \emph{transformations}) is applied for either high-level graph IRs and low-level Tensor IRs (TIR).
Within a pass sequence, each pass iteratively transforms an IR to a new IR to either optimize the computation or propagate valuable information for upcoming optimizations.
Once the low-level IR is ultimately optimized, the code generation component will produce corresponding binaries for different targets (i.e., NVIDIA GPU, X86 CPU, etc.)

Existing work on DL-library testing~\cite{lemon, cradle} mainly focuses on generation/mutation at the graph level. 
Contrastingly, for tensor compilers, we target the low-level representation since there are many limitations if the input files are simply constructed via such graph-level abstraction.
First, low-level IRs are closer to code generation and optimization which can guide the fuzzers to find deeper compiler bugs. 
Second, there is a limited search space for graph-level construction since deep learning operators are too coarse-grained and it suffers from various shape constraints. 
Furthermore, graph-level representation can be lowered to concrete low-level IR but not vice versa.
In this work, we have empirically compared our \sys technique that operates on the low-level IRs with state-of-the-art DL-library fuzzer, \lemon~\cite{lemon}, which performs graph-level model mutation. The evaluation results also confirm that \lemon generates 7.7x less valuable tests (i.e., the tests that are compilable and can trigger new compiler coverage) compared with \sys.

\subsection{Fuzzing} 
Fuzzing~\cite{afl,libfuzzer,lemieux2018fairfuzz,AFLplusplus-Woot20,bohme2017coverage}, known as an advanced automatic testing technique, has been widely employed to efficiently detect software bugs in the wild.
The key features of fuzzing, is the extreme 1) \emph{efficiency}: no heavy-weight analysis is required, and 2) \emph{simplicity}: fuzzers are mostly general-purpose and could be as easily employed as compiling a program and then executing it.

The big idea of fuzzing, is to generate randomized inputs in sharp and explore unexpected behaviors (e.g., crashes) of the program under test.
One of the most effective techniques of fuzzing is called the coverage-guided fuzzing (CGF), which is a \emph{mutation-based} approach that leverages coverage feedback to focus on test inputs (known as \emph{seeds}) that have achieved new coverage, instead of doing so in a randomized fashion.

The idea of CGF has led to many existing general-purpose binary fuzzers both in industry and in research~\cite{afl, libfuzzer, bohme2017coverage, AFLplusplus-Woot20, lemieux2018fairfuzz}.
AFL~\cite{afl} is one of the pioneers among CGF tools that have found numerous vulnerabilities in diverse applications. The development of AFL has inspired many further enhancements and extensions. 
AFLFast~\cite{bohme2017coverage} further leverages the Markov chain to model CGF as a systematic exploration of its state space and develops a set of power schedules and search strategies to focus on low-frequency paths. FairFuzz~\cite{lemieux2018fairfuzz}, which outperforms AFLFast in its evaluation, prioritizes seeds that hit rare branches, instead of rare paths, and develops a mutation mask algorithm to bias mutation towards producing inputs that hit such rare branches. AFL++~\cite{AFLplusplus-Woot20} further incorporates state-of-the-art fuzzing research ideas into one useful tool, which is prospective to be a new baseline tool for future research in Fuzzing.
LibFuzzer~\cite{libfuzzer}
has been widely recognized as one of the most representative coverage-guided fuzzers that builds in-process fuzzing loop and powerful evolutionary fuzzing engine with its integration with the LLVM infrastructure~\cite{lattner2002llvm}.
It has been under active development and keeping adopting the most recent and influential research ideas~\cite{bohme2020boosting}.

In addition to general-purpose fuzzers, CGF has also inspired many domain-specific fuzzers. \textsc{Die}~\cite{park2020fuzzing}, an aspect-preserving evolutionary fuzzing technique for JavaScript, has been shown to outperform state-of-the-art JavaScript fuzzers in terms of both bug discovery and valid test input generation. \textsc{Squirrel}~\cite{squirrel} is a database management system (DBMS) fuzzer that takes language validity into consideration during fuzzing, which has found numerous bugs in DBMSs including SQLite, MySQL, PostgreSQL, and MariaDB. FuzzChick~\cite{lampropoulos2019coverage}, an extension of QuickChick~\cite{denes2014quickchick}, incorporates coverage guidance to perform property-based testing for Coq programs, and has been shown to perform far better than the vanilla QuickChick with the help of coverage guidance.

The existing general-purpose fuzzers cannot be simply applied here for tensor compilers like TVM because tensor compilers require structural IRs in specific form as input, which does not have a direct correspondence to the binary stream.
\RevAdd{{(CHANGE LIST 2)}}{Furthermore, many traditional compiler fuzzing techniques~\cite{yang2011finding, le2014compiler, zhang2017skeletal}, though also theoretically general and applicable, are insufficient for tensor compiler fuzzing as they are not tailored for such purposes. For instance, the well-known EMI~\cite{le2014compiler} is general for any compilers supporting control flows. However, it is not suitable for DL computation as most existing DL models are static graphs (i.e., no control flows) mainly except for some RNN models. In addition, in TVM the de facto compilation mode (i.e., the ``graph'' mode) requires constant input tensor shape so that any control flows related to shape sizes can be statically inferred to allow maximum optimization (e.g., unrolling loops in an optimal way), making it unsuitable for applying EMI.}
To date, there are very few domain-specific fuzzers for tensor compilers, with TVMFuzz~\cite{tvmfuzz} being the only existing fuzzer specifically targeting TVM to our knowledge.
Therefore, this paper aims to build a practical fuzzing technique specifically targeting modern tensor compilers.%

\section{Approach}

\begin{figure}[h]
    \centering
    \includegraphics[width=\textwidth]{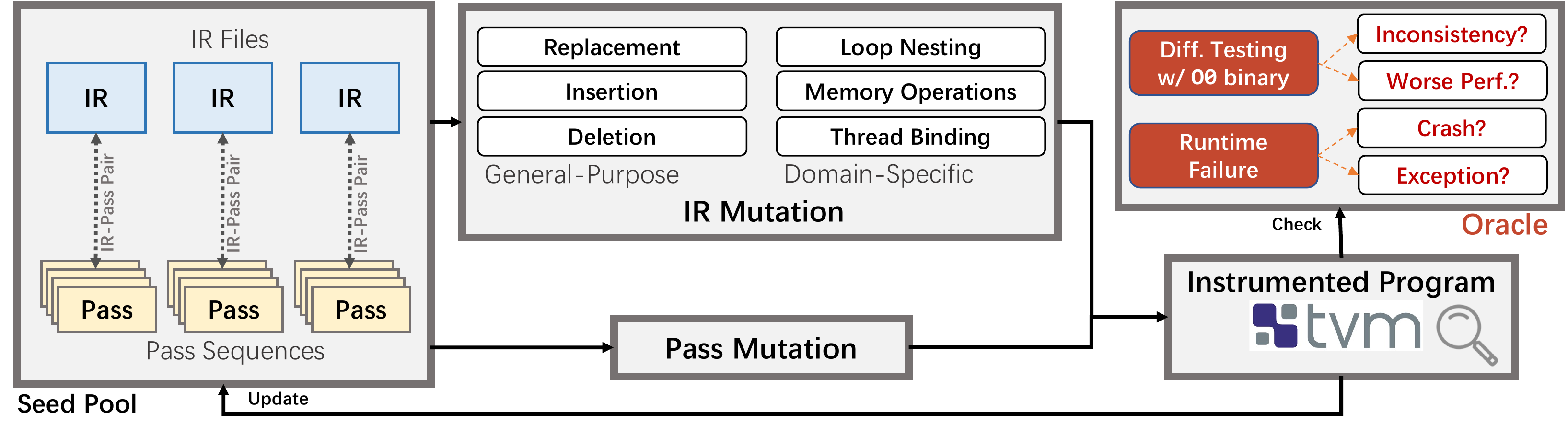}
    \caption{Overview of \sys{}}
    \label{fig:overview}
\end{figure}

In this section, we present the detailed design of \sys, a practical tensor compiler fuzzer via coverage-guided joint IR-Pass mutation. 
Figure~\ref{fig:overview} illustrates the overview of \sys. As shown in the figure, like traditional coverage-guided fuzzing work~\cite{li2018fuzzing}, \sys maintains a seed pool to store interesting seeds (i.e., the test inputs that can trigger new coverage) for further mutations. Different from prior work that mainly maintains the input files within the seed pool, 
\sys maintains two dimensions of information in the seed pool (i.e., both IR files and their corresponding optimization pass sequences) for effective joint IR-pass mutation.

During the fuzzing process, for each pair of IR and pass sequence from the seed pool, \sys{} will apply the corresponding mutation strategies to generate a new input pair in each iteration. For example, \sys{} applies both general-purpose and tensor-compiler-specific mutators on IR files to generate new IR files, and applies pass mutation to randomly generate a new pass sequence. Then, for each newly generated IR-pass pair, \sys leverages the tensor compiler under test (i.e., TVM in this work) to compile IR with the corresponding pass sequence and collect the compiler coverage information. Any input pairs that violate the test oracles are reported, while any input pairs that can help trigger new compiler coverage are further fed back to the seed pool for generating more valuable inputs. In this way, the generated inputs can cover more and more code for tensor compilers, and can detect more and more potential bugs. The fuzzing loop will terminate until the allowed time/resource budget runs out.

In the remainder of this section, we will first present the detailed algorithm design for our fuzzing loop (\S~\ref{sec:fuzzloop}). Then, we will present the details for our general-purpose mutators (\S~\ref{subsec:general-purpose-mutation}) and tensor-compiler-specific mutators (\S~\ref{sec:pass-spec}). Finally, we will briefly discuss the test oracle information used in this work (\S~\ref{sec:oracle}).  

\subsection{Fuzzing Loop}\label{sec:fuzzloop}

\begin{algorithm}[b]
\small
\caption{\sys Fuzzing Loop \MyBox[-1.5ex]{starta}{enda}{gray}\MyBoxH[-1.4ex]{startb}{endb}{gray}}\label{algo:fuzzloop}
\DontPrintSemicolon
\SetKwProg{Fn}{Function}{:}{}

\SetKwFunction{Fuzz}{\textsc{Fuzz}}
\Fn{\Fuzz{set of initial seeds $S_0$, time budget $\Time$, pass mutation frequency control $\NFail_{max}$}}{
$\Pool\leftarrow S_0$\;\label{algo:fuzzloop:init}
$C_{total}\leftarrow \bigcup_{i\in S_0} \textsc{Coverage}(i)$\;\label{algo:fuzzloop:collect}
\While{within time budget $\Time$\label{algo:fuzzloop:loopbeg}} {
    $\langle\File, \PassSeq, \NFail\rangle \leftarrow \textsc{Select}(\Pool)$\;
    \tikzmark{starta}\If{$\NFail = \NFail_{max}$\label{algo:fuzzloop:nmax}} {
        $\PassSeq'\leftarrow\textsc{MutatePass}(\PassSeq)$\;\label{algo:fuzzloop:pmut}
        $err, cov \leftarrow$ \RunTVM($\File, \PassSeq'$)\;\label{algo:fuzzloop:runtvm}
        \uIf{$\exists err$} {
            \textsc{Report}($\File, \PassSeq'$)\;\label{algo:fuzzloop:report}
        } \uElseIf{$cov\nsubseteq C_{total}$}{
            $C_{total}\leftarrow C_{total} \cup cov$\; \label{algo:fuzzloop:cov-update0}
            \textsc{Update}($\Pool,\langle\File,\PassSeq',0\rangle$)\;\label{algo:fuzzloop:update0}
        } \Else{
            \textsc{Update}($\Pool,\langle\File,\PassSeq,0\rangle$)\;\label{algo:fuzzloop:update1}
        }
        \textsc{Continue}\ \ \ \ \ \ \ \ \ \ \ \ \ \ \ \ \ \ \ \ \ \ \ \ \ \ \ \ \ \ \ \ \ \ \tikzmark{enda}
    }
    $\File'\leftarrow\textsc{MutateIR}(\File)$\;\label{algo:fuzzloop:fmut}
    $err, cov \leftarrow$ \RunTVM($\File', \PassSeq$)\;\label{algo:fuzzloop:compile1}
    \uIf{$\exists err$} {
        \textsc{Report}($\File', \PassSeq$)\;\label{algo:fuzzloop:report2}
    }\uElseIf{$cov\nsubseteq C_{total}$}{
        $\Pool\leftarrow\Pool\cup\langle\File',\PassSeq,0\rangle$\;\label{algo:fuzzloop:insert}
        $C_{total}\leftarrow C_{total} \cup cov$\; \label{algo:fuzzloop:cov-update1}
        \tikzmark{startb}\textsc{Update}($\Pool,\langle\File,\PassSeq,0\rangle$)\label{algo:fuzzloop:update2}
    } \Else{
        \textsc{Update}($\Pool,\langle\File,\PassSeq,\NFail + 1\rangle$)\label{algo:fuzzloop:update3}\ \ \ \ \ \ \ \ \ \ \ \ \tikzmark{endb}
    }
}\label{algo:fuzzloop:loopend}
}
\end{algorithm}

Algorithm~\ref{algo:fuzzloop} presents the detailed design of our main \sys fuzzing loop. The algorithm only takes three inputs, including the initial seed pool ($S_0$), the time budget ($T$), and the parameter for controlling the interleaving of IR and pass mutations ($\NFail_{max}$). 
Different from all prior work on evolutionary coverage-guided fuzzing~\cite{afl, libfuzzer}, the seed pool of \sys maintains two dimensions of information for effective tensor-compiler fuzzing, i.e., both the IR files and their corresponding pass sequences. Thus, we can denote each input for \sys as a pair $\langle\File, \PassSeq\rangle$, where $\File$ represents an IR file while $\PassSeq$ represents the corresponding pass sequence for the IR. In the algorithm, we further extend $\langle\File, \PassSeq\rangle$ into $\langle\File, \PassSeq, \NFail\rangle$ to additionally consider the interleaving control $\NFail$ for the join IR-pass mutation. With the interleaving control $\NFail$, for each seed input IR file $\File$, \sys can 1) keep mutating $\File$ with $\PassSeq$ if such mutations were rewarding, and 2) also occasionally (controlled by $\NFail$) seek a better $\PassSeq'$ to pair with $\File$ when the current $\PassSeq$ get stuck in local minima.

The main algorithm of \sys is similar to traditional evolutionary fuzzers, except for the additional code logic added to handle the additional pass mutation (highlighted in colored boxes). Basically, \sys first initializes the seed pool with pairs of IR files and pass sequences, as well as setting $\NFail$=0 for all pairs (Line~\ref{algo:fuzzloop:init}). For example, in this work, the initial seed pool consists of all possible model architectures in the TVM model zoo~\cite{tvm_model_zoo} with randomly generated pass sequences. The coverage achieved by the initial seed inputs is also collected to evaluate newly generated inputs (Line~\ref{algo:fuzzloop:collect}). Then, \sys will go through the main loop for generating new inputs (Lines~\ref{algo:fuzzloop:loopbeg}-\ref{algo:fuzzloop:loopend}). 

In each iteration, \sys will randomly fetch an input tuple from the seed pool. If the current $\langle\File, \PassSeq\rangle$ pair cannot trigger any new coverage during the past $\NFail_{max}$ consecutive IR mutations (Line~\ref{algo:fuzzloop:nmax}), \sys will try to mutate the pass sequence $\PassSeq$ into another random sequence $\PassSeq'$ in the hope that $\PassSeq'$ will bring this input pair to a better state for further mutations (Line~\ref{algo:fuzzloop:pmut}). The coverage and error information will be recorded when compiling the input pair $\langle\File, \PassSeq'\rangle$ with the compiler under test (Line~\ref{algo:fuzzloop:runtvm}). In case of any error, the input pair will be reported to the developers. 
If $\PassSeq'$ does help trigger new coverage, the total coverage information will be updated (Line~\ref{algo:fuzzloop:cov-update0}); the input pair $\langle\File, \PassSeq, \NFail\rangle$ in the seed pool will also be updated to $\langle\File, \PassSeq', 0\rangle$ since it is more promising to go with $\PassSeq'$ in future runs on mutating $\File$ (Lines~\ref{algo:fuzzloop:update0}). 
If $\PassSeq'$ does not help trigger new coverage, \sys simply clears the interleaving control counter to 0 to allow more file mutations with the current $\PassSeq$ (Line~\ref{algo:fuzzloop:update1}). This indicates that 
we do not perform consecutive pass mutations for any seed input regardless of the coverage outcome. The reason is that mutating pass sequences is not as rewarding as mutating IR files in general, and we only need \emph{occasional} pass mutation (controlled via $\NFail$) to guide the evolutionary process to more promising states to avoid local minima.

On the other hand, if the fetched input tuple has not failed to trigger new coverage for $\NFail_{max}$ consecutive IR mutations, \sys will go ahead to further mutate the IR file following a very similar process to traditional fuzzers. 
\sys first mutates the IR $\File$ into $\File'$ by selecting one mutator among the mutator pool (including 3 general-purpose and \NPSpecMut{} \pspec{} mutators), and then collects the result information for compiling the pair $\langle\File', \PassSeq\rangle$ (Lines~\ref{algo:fuzzloop:fmut} and \ref{algo:fuzzloop:compile1}).
In case of any error, the input pair will be reported. If $\File'$ does help trigger new coverage, the new IR file with the current $\PassSeq$ will be inserted into the seed pool for future runs (Line~\ref{algo:fuzzloop:insert}). The total coverage information will also be updated (Line~\ref{algo:fuzzloop:cov-update1}). Different from prior fuzzers, \sys also needs to update the original seed pair to $\langle\File, \PassSeq, 0\rangle$ since it helped trigger new coverage (Line~\ref{algo:fuzzloop:update2}); also, if $F'$ did not help achieve new coverage, the original seed pair will be updated to $\langle\File, \PassSeq, \NFail+1\rangle$ to record the current attempt that failed to trigger new coverage (Line~\ref{algo:fuzzloop:update3}). 

\RevAdd{(CHANGE LIST 4)}{Theoretically, some specific $\langle\File, \PassSeq\rangle$ might fail due to 1) lack of pass dependency, or 2) pass/IR incompatibility, resulting in waste of time for compiling invalid $\langle\File, \PassSeq\rangle$. Executing too many invalid compilations will make fuzzing process less efficient. The evolutionary joint IR-Pass mutation (Algorithm~\ref{algo:fuzzloop}) can easily avoid such frequent invalid compilation by design. 
As is shown in Line~\ref{algo:fuzzloop:insert}, only valid $\langle\File, \PassSeq\rangle$ with new coverage will be added into the seed pool $\Pool$, whereas the invalid and ineffective ones will be ignored to keep seed pool filled with compilable samples during evolutionary fuzzing process.}

In this way, after being launched, the algorithm can then continuously generate valuable IR and pass sequence pairs for triggering tensor-compiler bugs.

\subsection{General-Purpose Mutation}
\label{subsec:general-purpose-mutation}

Following prior work on fuzzing programming languages~\cite{squirrel, holler2012fuzzing, lampropoulos2019coverage}, we design a general-purpose IR mutation approach.%
In addition, program analysis techniques are integrated into mutation to ensure syntax correctness and to mitigate semantic errors.
This is because, to dig high-quality bugs in the code generation and optimization phases of a compiler,
the produced IRs should be able to pass standard pre-condition checks (e.g., syntax checks and semantics checks). 
In the remainder of this section, we first introduce our definition of the low-level Tensor IR (TIR) of TVM%
and then elaborate on the the mutation details.

\begin{figure}
    \centering
    \includegraphics[width=\textwidth, 
    page=1]{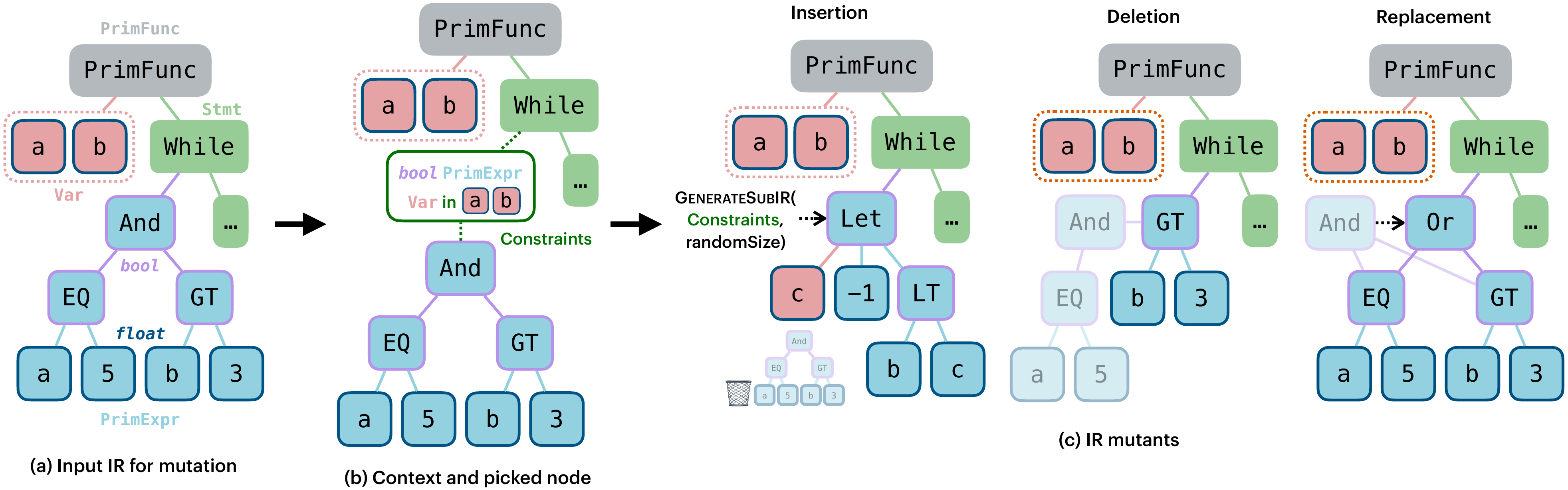}
    \caption{The process of IR mutation. 
    Node types are differentiated using background colors, while expression types
    are differentiated using border colors. Each label around the node denotes a node type or an expression type. Each label on the node denotes a constructor or a primitive. The \cons{VarInjection} constructor in Table~\ref{tab:tir-ast} is illustrated by switching the background color of the \cons{Var} node.}
    \label{fig:mut-example}
\end{figure}

We first discuss the abstract syntax tree (AST) of TIR since it is the entry point of TVM's compilation.
Figure~\ref{fig:mut-example}a depicts a simplified TIR AST sample.
The AST tree contains different types of nodes, with the root node representing the input IR to the compiler.
As is shown Figure~\ref{fig:mut-example}a, the type of the root node is \type{PrimFunc} which stands for the basic function type in TIR.
The \cons{While} node is of type \type{Stmt} while the \cons{EQ} and \cons{GT} nodes are of type \type{PrimExpr}.
The corpus of all node types can be defined as \irTypeCorpus{}%
We assume these types are disjoint (e.g., no subtype relation), but our implementation leverages the concept of \Constructor to simulate the original subtype relations.
For each type, there could be multiple \emph{constructors}, which are functions/operators from one or more node types to one return node type, generally having the signature \irConstructorDefinition{}%
Some types also have \emph{primitives}, which are values that cannot be broken down into subparts (i.e., leaf nodes).%

Table~\ref{tab:tir-ast} shows a detailed list of common TIR AST node types, constructors, and primitives.
Note that \cons{VarInjection}, one \Constructor of \type{PrimExpr}, is added by us to switch the variable from type \type{Var} to \type{PrimExpr} without changing its internal value.%
This is required because \type{Var} is a \emph{subtype} of \type{PrimExpr} in the implementation of TIR by TVM, which means each \type{Var} is implicitly a \type{PrimExpr}, but in our definition, we assume no subtype relation. By using injective constructors, this could be easily expressed. 

\begin{table}[htbp]
    \footnotesize%
    \centering
    \resizebox{\textwidth}{!}{%
    \begin{tabular}{p{4cm}l}
        \toprule
        Node Type 
        & Constructors and Primitives\\
        \midrule
        
        $\type{PrimFunc.}$ Function type, the input type of the compiler.
            & $\begin{aligned}[t]
                &\cons{PrimFunc}: (\type{Var}^*, \type{Stmt}) \to \type{PrimFunc}\\
                &\cons{PrimFunc}: (\type{Var}^*, \type{Stmt}, \type{Buffer}) \to \type{PrimFunc}\\
              \end{aligned}$\\
        \midrule
        
        $\type{Buffer.}$ Buffer type, describing the storage of data.
            & $\begin{aligned}[t]
                &\cons{decl\_buffer}: (\type{Shape}, \type{DataType}) \to \type{Buffer}\\
              \end{aligned}$\\
        \midrule
        
        $\type{DataType.}$ Basic numeric data types for variables and expressions, including integer, boolean, floating point, etc.
            & $\begin{aligned}[t]
                &\cons{float32}: \type{DataType}\\
                &\cons{int32}: \type{DataType}\\
                &\cons{uint32}: \type{DataType}\\
                &\cons{bool}: \type{DataType}\\
              \end{aligned}$\\
        \midrule
              
        $\type{Var.}$ Variable type, used for variable declaration and as expressions.
            & $\begin{aligned}[t]
                &\cons{Var}: (\mathit{name}: \type{String}, \type{DataType}) \to \type{Var}          
              \end{aligned}$\\
         \midrule
         
         $\type{Stmt.}$ Statement type, the fundamental building block to form a function. 
         There are sequential statements, control flow statements, etc. 
         Part\-i\-c\-u\-larly, statements constructed by the \cons{For} constructor are central to many low level optimizations.
            & $\begin{aligned}[t]
                &\cons{While}: (\type{PrimExpr}, \type{Stmt}) \to \type{Stmt}\\
                &\cons{For}: (\type{Var}, \mathit{min}: \type{PrimExpr}, \mathit{extent}: \type{PrimExpr}, \type{ForKind}, \type{Stmt}) \to \type{Stmt}\\
                &\cons{IfThenElse}: (\type{PrimExpr}, \type{Stmt}, \type{Stmt}) \to \type{Stmt}\\
                &\cons{LetStmt}: (\type{Var}, \type{PrimExpr}, \type{Stmt}) \to \type{Stmt}\\
                &\cons{SeqStmt}: (\type{Stmt}^*) \to \type{Stmt}\\
                &\cons{BufferStore}: (\type{Buffer}, \type{PrimExpr}, \mathit{indices}: \type{PrimExpr}^*) \to \type{Stmt}\\
               \end{aligned}$\\
         \midrule
         
         $\type{PrimExpr.}$ %
         Expression type, the fundamental building block to form a statement. The constructors include basic operators that handle numeric values, buffers, etc. The \cons{Call} constructor is responsible for constructing pre-defined \emph{intrinsics} by TVM. Each \type{PrimExpr} has a corresponding \type{DataType}, either specified explicitly or inferred implicitly.
            & $\begin{aligned}[t]
                &\cons{VarInjection}: (\type{Var}) \to \type{PrimExpr}\\
                &\cons{And}: (\type{PrimExpr}, \type{PrimExpr}) \to \type{PrimExpr}\\
                &\cons{Or}: (\type{PrimExpr}, \type{PrimExpr}) \to \type{PrimExpr}\\
                &\cons{EQ}: (\type{PrimExpr}, \type{PrimExpr}) \to \type{PrimExpr}\\
                &\cons{GT}: (\type{PrimExpr}, \type{PrimExpr}) \to \type{PrimExpr}\\
                &\cons{LT}: (\type{PrimExpr}, \type{PrimExpr}) \to \type{PrimExpr}\\
                &\cons{Add}: (\type{PrimExpr}, \type{PrimExpr}) \to \type{PrimExpr}\\
                &\cons{Call}: (\type{DataType}, \type{Op}, \mathit{args}: \type{PrimExpr}^*) \to \type{PrimExpr}\\
                &\cons{Let}: (\type{Var}, \type{PrimExpr}, \type{PrimExpr}) \to \type{PrimExpr}\\
                &\cons{BufferLoad}: (\type{Buffer}, \mathit{indices}: \type{PrimExpr}^*) \to \type{PrimExpr}\\
                &\cons{FloatImm}: (\type{DataType}, \type{Float}) \to \type{PrimExpr}\\
                &\cons{IntImm}: (\type{DataType}, \type{Int}) \to \type{PrimExpr}\\
               \end{aligned}$\\
        \bottomrule
    \end{tabular}}
    \caption{Example node types, constructors, and primitives of the AST of TIR for Figure~\ref{fig:mut-example}. Some constructor could be overloaded or have the same name as their node types. The asterisk `*' means a list type  (e.g., $\mathtt{Var}^*$ in the two $\mathtt{PrimFunc}$ constructors means a list of $\mathtt{Var}$, which serves as parameters of a function). We also put several auxiliary labels in front of some parameter types to help understand the meaning of the parameter (e.g., ``$\mathit{name}: \mathtt{String}$'' in the $\mathtt{Var}$ constructor signature). 
    }
    \label{tab:tir-ast}
\end{table}

Each AST node can be recursively defined as either a primitive (leaf node) or an application of some constructor to other nodes (e.g., branch node).
For simplicity, in our implementation of mutation approaches, some trivial branch nodes are treated as leaf nodes, including \cons{Var}, \cons{IntImm}, \cons{FloatImm}, etc.%
As an example, the root node of the input IR in Figure~\ref{fig:mut-example}a can be formally defined as \irExampleDefinition{}%

The first step of our mutation approach is to randomly pick out one of the AST nodes of the given IR and regard it as a hole, which can then be filled up to produce an IR mutant. We call an IR with a hole at some position a \emph{context}.
For instance, in Figure~\ref{fig:mut-example}b,
we pick out the \cons{And} node as a hole (denoted by $\square{}$) so that the corresponding context is \irExampleContext{}%
Based on the context, we can derive the \emph{constraints} to be satisfied (e.g., accessible variables of the hole) when filling the hole so that the filled IR could be correct.
Formally, the constraints are a tuple of necessary information that helps determine the requirements when constructing a sub-expression in the hole. \RevAdd{(CHANGE LIST 6)}{Specifically, for TIR, we consider the following information:
\begin{itemize}
    \item \emph{Desired AST node type (e.g., \type{PrimExpr}, \type{Stmt}, \type{Var}).}
    \item \emph{Desired expression type (e.g., \cons{int32}, \cons{float32}, \cons{bool}).}
    \item \emph{Accessible variables under the current scope.}
    \item \emph{Declared buffers.} TIR uses the notion of “buffer” to store and load data. When we access a buffer, we should ensure it is already declared.
    \item \emph{A boolean indicating whether the variables need to be bound.} TIR only allows a commented expression to have free variables.
\end{itemize}}
As an example, for the context in Equation~\ref{eq:irExampleContext}, the hole represents a condition check for the \cons{While} node. Hence, in order to fill the hole, at least a boolean expression is needed. Also, any variable used should be bound to some binding occurrence (e.g., parameter \cons{a} and \cons{b}). Therefore, the constraints should be \irExampleConstraints{}%
Based on the derived constraints and the picked node, we perform a series of \emph{mutations} using the corresponding \emph{mutator} on the node following the constraints. Formally, each mutator has the signature \irMutationDefinition{}%
We use a disjoint union here because our mutator is designed to operate on nodes of any node type and different node types should not overlap with each other in our definition.

Basically, we designed the following three general-purpose mutators, namely \emph{Insertion}, \emph{Deletion}, and \emph{Replacement}:

\parabf{Insertion.} Regardless of the input node, \sys{} simply returns a new node generated from scratch that satisfies the given constraints. This is done by \sys{}'s generator, which is inspired by the prior generators in the random testing community~\cite{claessen2015generating, lampropoulos2017generating}.
The functionality of the generator is to produce IR ingredients/snippets based on the constraints and a \emph{size} parameter which indicates the node size of generated sub-IR, as is described in Figure~\ref{fig:mut-example}c.%
In the figure, \sys{} generates a new boolean \cons{Let} node of type \type{PrimExpr}, and ensures that all the variable references have their corresponding binding occurrences (e.g., in the node $\cons{LT}(\cons b, \cons c)$, \cons{b} is introduced by the parameter list, and \cons{c} is introduced by \cons{Let}).

\parabf{Deletion.} \sys{} checks the child nodes of the input node, filters out those satisfying the constraints, and randomly returns one of them. For example, in Figure~\ref{fig:mut-example}c, we perform deletion on the \cons{And} node by returning its right-hand side $\cons{GT}(\cons b, \cons 3)$, the `greater than' node, which is a boolean expression with all variable references bound.

\parabf{Replacement.} For a primitive node, \sys{} simply modifies its value, or returns another primitive based on the constraints. For a node constructed by some constructor, in the simplest case, \sys{} randomly selects a constructor to substitute the existing one, in the restriction that after the substitution the node should satisfy the constraints given. More generally, \sys{} randomly selects a constructor, trying to use the child nodes of the input node as components to fill the parameter list of the selected constructor; if there are parameters unable to fill, \sys{} randomly generates one using the generator. This strategy is inspired by the $\mathtt{mutate}_g^{T}$ constructor of FuzzChick~\cite{lampropoulos2019coverage} for testing Coq programs except that \sys{} considers different constraints. Figure~\ref{fig:mut-example}c gives the simplest form of replacement, which just replaces the \cons{And} constructor with the \cons{Or} constructor.

\subsection{Domain-Specific Mutation}
\label{sec:pass-spec}

Tensor compilers
focus on optimizing domain-specific programs, e.g., programs with dense loops in particular.
To optimize those hot spot program structures, 
existing tensor compilers~\cite{tvm,plaidml,xla,jin2020compiling} leverage the concept of \emph{pass} to optimize the given IR or insert annotations containing valuable information for further optimization.
To trigger the complex logic behind those optimization passes, general-purpose mutators, though versatile to handle different types of expressions, are still inefficient and not tailored to the specific domain that tensor compilers are built for.

For domain-specific compiler testing, in addition to the general-purpose mutators, we argue that it is also important to navigate the mutation towards the core components that the compilers specifically target (e.g., loop-oriented optimization, memory allocation, memory latency hiding, and parallelization~\cite{li2020deep}).
For example, deep and wide nested loops can be optimized with tiling~\cite{tiling}, multi-threading~\cite{threading}, and vectorization~\cite{vectorization} by a series of related passes (e.g., \texttt{UnrollLoop} and \texttt{LoopPartition}).
Those passes have complex optimization rules for different domain-specific code structures (e.g., big loops, large buffer allocation, and thread scheduling) that general-purpose mutators can hardly target.
Hence, according to the hot spot program patterns targeted by existing tensor compilers~\cite{li2020deep,tvm,halide,zhao2021akg,tillet2019triton}, \sys specifically designed 3 types of mutators:
1) \emph{loop-nesting} mutator for creating multifarious dense loop structures; 
2) \emph{memory-operation} mutator for various memory allocation/store/load patterns at the index level; and
3) \emph{thread-binding} mutator for diversifying the parallel computation flows to generate interesting code patterns that tensor compilers particularly care about.

\noindent\textbf{Loop Nesting.}
Tensor computation usually consists of a large number of nested loops.
Even for the simplest element-wise expression, e.g., \texttt{C=C+1} with broadcasting,
the loop structure of a common image tensor (whose dimensions are [height, width, channels]) will consist of 3 nested loops.
To mimic such dense loops, 
we introduce the loop-nesting mutator to transform IRs with different loop structures. %

\begin{figure}
    \centering
    \includegraphics[width=\textwidth]{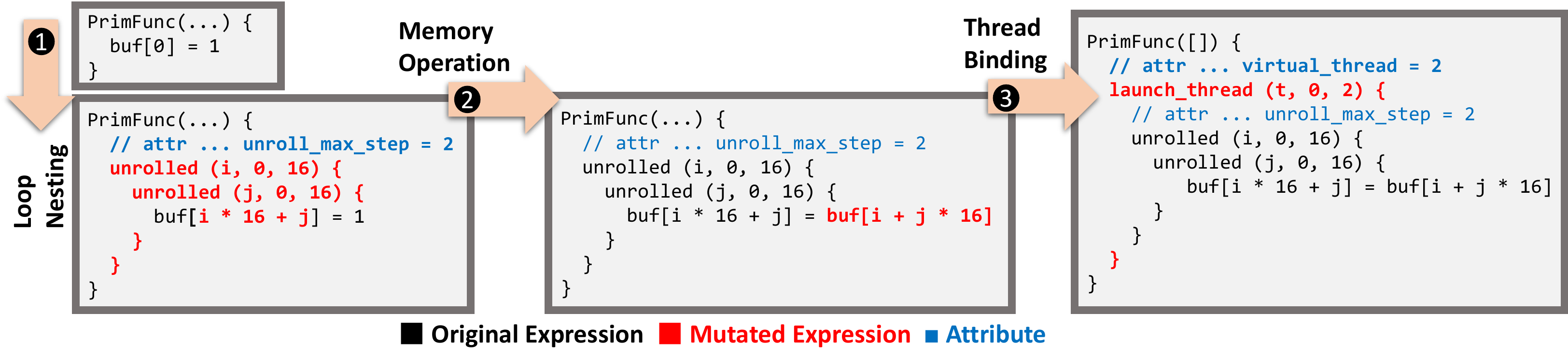}
    \caption{Example of \PSpec{} IR Mutation}
    \label{fig:pass-specific-example}
\end{figure}

First, \sys randomly picks an AST node as the innermost loop body.
\sys then selects one out of the five TVM loop types (\texttt{serial}, \texttt{vectorize}, \texttt{unroll}, etc.),
where each of them represents different control flow semantics.
Given the loop type, \sys inserts several loops with either constant or variable loop sizes (constant loops are likely to trigger loop unrolling while variable loops will block such optimization).
For example, in step \blackding{1} of Figure~\ref{fig:pass-specific-example}, 2 nested loops of type \texttt{unrolled} are inserted after mutation. 
Furthermore, according to loop variables under the current context, a random expression will be used to form the indices (\texttt{[i*16+j]}).
Notably, TVM also annotates loop attributes for concrete optimization in code generation. e.g., \texttt{unroll\_max\_steps}, and further tunes those integer attributes to trigger different optimization paths.
 Therefore, \sys also mutates those attributes when creating/replacing the target loops.

\noindent\textbf{Memory Operations.}
Apart from multifarious loop structures, another dimension to increasing the complexity of tensor computation is to introduce various memory operations, including memory store/load and allocation.

\sys's memory-operation mutator mimics complex memory patterns by inserting memory operations into existing IRs.
Given a randomly selected node
, \sys first analyzes accessible memory buffers (represented with pointers) under the current scope.
Next, \sys randomly constructs a memory operation (i.e., a sub-expression) and inserts it into the target AST node.
As is shown in Figure~\ref{fig:pass-specific-example} (step \blackding{2}), \sys inserts a sub-expression (i.e., \texttt{... = buf[i+j*16]}) to original IR so that a new memory access is created and the dataflow related to \texttt{buf} is changed.

\noindent\textbf{Thread Binding.}
One thing that differentiates traditional compilers and tensor compilers is that tensor compilers leverage multiple threads (either CPU threads or threads of parallel hardware like NVIDIA GPU) to automatically parallelize the program.
The thread scheduling, however, could have many different settings, as operations could be executed by different thread groups at different stages (manipulated by attributes, e.g., thread numbers and thread tags).

To explore the impact of different thread scheduling patterns, \sys creates various thread-binding patterns and leverages them to mutate the multi-thread planning of given IRs.
Precisely, as is shown in Figure~\ref{fig:pass-specific-example} (step \blackding{3}), \sys first selects an AST node (i.e., the 2 nested loops wrapped by the scope of \texttt{launch\_thread}) and then initializes its threading parameters, e.g., virtual thread number (\texttt{virtual\_thread} in TVM).
In this way, \texttt{virtual\_thread} is initialized by 2 which means this node will be executed by 2 virtual threads.

\subsection{Test Oracle}\label{sec:oracle}

Test oracles are important for detecting potential bugs with fuzzing. In this paper, we consider the following ways to resolve the test oracle problem for finding bugs in tensor compilers:

\parabf{Result Inconsistency.}
\sys holds the hypothesis that an IR, whether it is optimized or not, should keep the output result consistent.
For each generated IR, \sys will compile it twice, where it first compiles the IR with the lowest optimization and then compiles it with given optimization passes.
In this way, \sys compares the output results by feeding 2 model binaries the same input data.
We identify it as an inconsistency bug if the absolute or relative error exceeds the expectation.

\parabf{Performance Degradation.}
The second hypothesis by \sys is that after a series of optimization passes, the performance should not be degraded.
Therefore, \sys would instrument the running time of optimized and non-optimized executions.
If the optimized code runs even slower than the non-optimized one, we consider it as a potential performance bug.
\RevAdd{(CHANGE LIST 6)}{Notably, to avoid false-positives, we set clear performance margin in the differential testing setting. The non-optimized version is compiled with lowest optimization level (\texttt{opt\_level=0}) while the optimized one is compiled with highest optimization level (\texttt{opt\_level=4}). Note that higher optimization level allows better and more aggressive optimization than lower levels given the same pass sequences. For example, level-3 graph fusion (i.e., \texttt{FuseOps}) allows more operator fusion patterns than the low-level one.} 

\parabf{Crash and Unexpected Exception.}
Like most Python applications, throwing an exception is the default behavior of errors.
Hence, Python/C++ projects (e.g., most tensor compilers) need to convert C++ exceptions into Python ones.
For example, in TVM's C++ codebase, any unexpected behavior (e.g., assertion failure) will result in C++ exceptions, where the top-level foreign function interface (FFI) handler will catch such C++ exceptions and pack the error message using the type \texttt{TVMError} for Python front-end.
Therefore, though errors might occur, the symptom should be uncaught exceptions rather than crash.%
The compilation and execution phase of \sys is done by forking a sub-process, \sys observes such crash by checking the return code of sub-processes.
\sys{} also monitors exceptions thrown during compilation as potential bugs.
To avoid false alarms, \sys{} has made the best effort on constructing legal IRs and pass sequences.

\section{Experimental Setup}

\subsection{Research Questions}

In this paper, we study the following research questions to thoroughly evaluate \sys:

\begin{itemize}[leftmargin=*]
  \item \textbf{RQ1:} How is the effectiveness of \sys{} compared with state-of-the-art fuzzing techniques on testing the TVM tensor compiler?
  \item \textbf{RQ2:} Are all components of \sys{} contributing positive improvements to its final effectiveness?
  \item \textbf{RQ3:} How do different parameter settings and experimental setups affect \sys's effectiveness?
  \item \textbf{RQ4:} How effective is \sys{} in detecting previously unknown bugs?
\end{itemize}

\RevAdd{(CHANGE LIST 3)}{The consideration of our experiment design largely follows suggestions made by~\citet{klees2018evaluating}.
The main differences are caused by the fuzzing targets, i.e., \citet{klees2018evaluating} mainly studied binary fuzzing while we are working on tensor compiler fuzzing. For example, the paper suggested a 24-hour timeout, while we evaluate \sys with a default 4h timeout since existing techniques tend to saturate within 4 hour.
Meanwhile, we do evaluate \sys with a 24-hour budget as well in RQ3.}

\subsection{Implementation}

\sys has been mainly implemented in 8.7k lines of Python code and \cppLine lines of C++ code for coverage extension with the following main components:

\parabf{Mutators.}
    We implemented all the 3 general-purpose mutators and \NPSpecMut \pspec{} mutators via directly operating on TIR in-memory objects (i.e., \texttt{tir.PrimFunc}) for fast mutation.
    More specifically, the mutation procedure is implemented by extending the visitor pattern of TIR's recursive post-order traversal interface.
    In addition, the utility generator used by replacement and insertion is capable of constructing various sub-expressions based on 89 TIR operator APIs.
    When inserting/replacing sub-expressions into an existing TIR, we consider the syntactic/semantic correctness \RevAdd{(CHANGE LIST 6)}{by maintaining IR constraints during the visiting process (e.g., preventing the use of variables that are undeclared or out of the scope).
    We further utilize casting nodes when generating intrinsic function calls. Although casting is not necessary theoretically due to our constraints-based approach, TVM provides more than 30 intrinsics whose detailed function signatures may vary and are not documented (e.g., \texttt{tir.cos} returns \texttt{float} whereas \texttt{tir.clz} returns \texttt{int}). To save manual efforts, we simply regard those intrinsics as opaque ones and cast them to satisfy the constraints.}
    
\parabf{Executor.} Once \sys{} generates a TIR file and pass sequence pair, they are sent to a sub-process for compilation and execution. The sub-process mechanism is to provide process-level isolation so that the fuzzing loop continues even though the TIR file and pass sequence make the sub-process crash.
    
\parabf{Coverage Collector.} 
    We implemented \texttt{memcov}, our in-memory coverage instrumentation tool, by extending LLVM's Coverage Sanitizer (i.e., injecting a customized function when entering each of CFG edges in the target program). 
    Once a program is compiled along with \texttt{memcov}, 
    we maintain a bit vector whose size is exactly the number of CFG edges of the instrumented program (i.e., TVM).
    When entering one edge, its corresponding position on the bit vector is set to \texttt{True}.
    As we implemented \sys's core components in Python, we also provide a Python interface to get the coverage state at that point by invoking C++ functions through \texttt{ctypes}~\cite{ctypes} (a Python-C++ FFI tool).

\parabf{Reporter.} Once a test violates our test oracle, the reporter would record necessary contextual data to reproduce the failure and debugging.

Consistent with Algorithm~\ref{algo:fuzzloop}, the \sys implementation takes three inputs, i.e., $S_0$, $\Time$, and $\NFail_{max}$. For the initial seed pool $S_0$, by default \sys uses 629 TIR functions converted from all possible official models from TVM's model zoo (\texttt{tvm.relay.testing}); for the time budget $\Time$, by default \sys sets it to 4 hours; for the IR-pass mutation control $\NFail_{max}$, by default \sys sets it to 5. We use such default setting for \sys unless explicitly specified, e.g., we will present the detailed impacts of different parameter settings on \sys in RQ2 (\S~\ref{sec:exp:rq2}).

\RevAdd{(CHANGE LIST 1)}{The main techniques behind \sys are general to other tensor and even traditional compilers which model low-level IRs and optimization passes. To implement our approaches for a new compiler, one needs to implement language mutators following rules described in \S~\ref{subsec:general-purpose-mutation} and \S~\ref{sec:pass-spec}, as well as figuring out corresponding optimization passes. The syntactic and semantic correctness of mutated IRs and passes should also be maintained. After that, the main algorithm and skeleton of \sys shall directly apply.}

\subsection{Compared Work}

To faithfully evaluate the effectiveness of \sys, we compare \sys{} with both the state-of-the-art general-purpose fuzzers and domain-specific fuzzers that can be applied/adapted for TVM fuzzing. 
More specifically, we include the following representative techniques in our evaluation:

\begin{itemize}
    \item \tvmfuzz~\cite{tvmfuzz}: This is the only existing fuzzer specifically targeting TVM to our knowledge. It follows a pure generation-based approach, which randomly generates TIR expressions by crafting valid expression ASTs of TIR. The generation approach is based on a user-defined probability table for different TIR nodes, while the validity is achieved by casting the input expressions to the parameter types of the operator.
    \item LibFuzzer~\cite{libfuzzer}: This is one of the state-of-the-art bit-level general-purpose binary fuzzers.
    It has been adopted as the first fuzzer supported by the famous Google OSS-Fuzz project~\cite{serebryany2017oss}, which has found thousands of security vulnerabilities and stability bugs; furthermore, it is also the officially used fuzzer for many popular projects including Chrome~\cite{libfuzz-chrome} and glibc~\cite{libfuzz-glibc}. 
     In this work, for a fair comparison with \sys, we also run LibFuzzer with the TVM official model files (exported in JSON) as seeds for fuzzing TVM. 
    \item \lemon~\cite{lemon}: \lemon is the state-of-the-art graph-level model generator for testing the operator-level DL libraries. 
    At the graph level, different operators in a computation graph usually have various tensor shape constraints that are very complex to resolve.
    To resolve this difficulty, \lemon developed a series of mutators for shape-invariant operators and their compositions, by replacing operators with equivalent shape requirements or inserting/deleting element-wise operators.
    Since \lemon mutates the high-level computation graphs, its generated models can be directly applied to simulate TVM fuzzing at the high level. For a more fair comparison with \lemon, we also run a \sys variant with \lemon's model seeds (this is because \lemon leverages Keras~\cite{keras} model files which can be converted to TIR but cannot be done vice versa).
\end{itemize}

\subsection{Metrics}\label{sec:metric}
We use the following metrics to evaluate the performance of \sys and the compared techniques:

\noindent\textbf{Code Coverage.}
Code coverage has been widely recognized as one of the most widely used metrics to evaluate software testing techniques~\cite{gopinath2014code}. The reason is that it is impossible for testing techniques to detect bugs in a code portion without actually executing it. Surprisingly, although existing work on testing deep learning libraries~\cite{cradle, lemon} claimed to cover more library code, they failed to present the detailed code coverage information. In this work, we instrument the entire TVM code base by extending LLVM's Coverage Sanitizer
and collect the detailed code coverage information at the edge level
for the studied techniques to thoroughly evaluate their test effectiveness.
Note that since we are comparing techniques for fuzzing the TVM compilation process, to make the comparison fair, we omit the coverage brought by other irrelevant modules at the initialization phase (e.g., constructing TIR functions by converting input models).

\noindent\textbf{Number of Valuable Tests.} 
Following prior work on fuzzing~\cite{park2020fuzzing}, for each compared technique, we also present the number of generated \emph{valuable} tests, i.e., the tests that are not only valid (i.e., compilable) but also contribute new coverage during the fuzzing process.
This metric is essential since the number of syntactically/semantically valid tests with new coverage can largely indicate the number of unique system behaviors/paths covered/tested. Also, this metric can largely complement code coverage, because techniques that mostly generate invalid inputs can still achieve high coverage for the error-handling code but that is clearly not what we want.

\noindent\textbf{Number of Detected Bugs.}\label{sec:bug_count}
Following almost all prior work on software testing and fuzzing~\cite{manes2019art,li2018fuzzing}, we further present the number of previously unknown bugs detected by all the studied techniques since bug detection is the ultimate goal for such techniques. In this work, we distinguish different bugs based on how they are fundamentally fixed. For instance, we found that 21 TIR operator functions (such as \texttt{tir.op.clz(None)}) will crash when given NULL inputs on a specific TVM version, but we only count this as 1 bug since all the crashes can be fixed by changing only one C++ macro statement.  %

\subsection{Experimental Procedure}\label{sec:exp:procedure}

For a fair comparison, we collect coverage of all compared techniques with the default 4-hour time budget using the same in-memory coverage collector that we implemented based on LLVM Coverage Sanitizer. Note that for \tvmfuzz and other baselines requiring no coverage feedback, we first run them on non-instrumented TVM binary for 4 hours to prevent unnecessary overhead introduced by coverage tracing. 
Then, we collect the generated TIR files and passes (if any) from them, and run compilation for them on instrumented TVM binary for offline coverage analysis.
Notably, for \lemon, we collect the Keras~\cite{keras} models generated in 4 hours, and convert them to TIR functions.
We then run the TIR functions on instrumented TVM to mimic the effectiveness of \lemon's graph-level construction for fuzzing TVM.
Of course, for those studied techniques requiring coverage feedback, we directly record the coverage within one run on instrumented TVM.

We conducted experiments on: 1) \emph{GPU test-bed}: a test-bed with Intel i9-9900X CPU (10 physical cores), GeForce RTX 2080 Ti GPU, and 128GB RAM, running 64-bit Ubuntu 18.04 as the operating system; and 
2) \emph{CPU test-bed}: a virtual cloud server (Alibaba Cloud ecs.c6e instance) with 4 CPU cores and 8GB RAM, running 64-bit Ubuntu 20.04.
Since one of the baselines, \lemon, requires a GPU environment, we did RQ1 (comparison with existing work) on the \emph{GPU test-bed} and all other RQs on the \emph{CPU test-bed}.
To ensure performance fairness, we made the system environment \RevDel{as pure as possible}\RevAdd{}{exclusive to the benchmarks} so that the system average load is always around 1 during the process.
For instrumentation, we compiled TVM v0.8-dev (\texttt{9b034d7}) with LLVM-12 and leveraged Coverage Sanitizer to trace edge coverage.
TVM is compiled under optimization level \texttt{O2} and other configurations are set by the default value.
\RevAdd{(CHANGE LIST 5)}{Since TVM contains as many as 17 targets, 4 executors, and many other irrelevant utilities (e.g., debuggers and profilers), in our evaluation, we focused on the LLVM-X86 target and the graph executor as they are widely adopted in TVM’s tutorials and in practice.}
\section{Result Analysis}

\subsection{RQ1: Comparison with Existing Work}\label{sec:exp:rq1}

\begin{figure}[h]
    \centering
\begin{minipage}{.48\textwidth}
    \includegraphics[width=\textwidth]{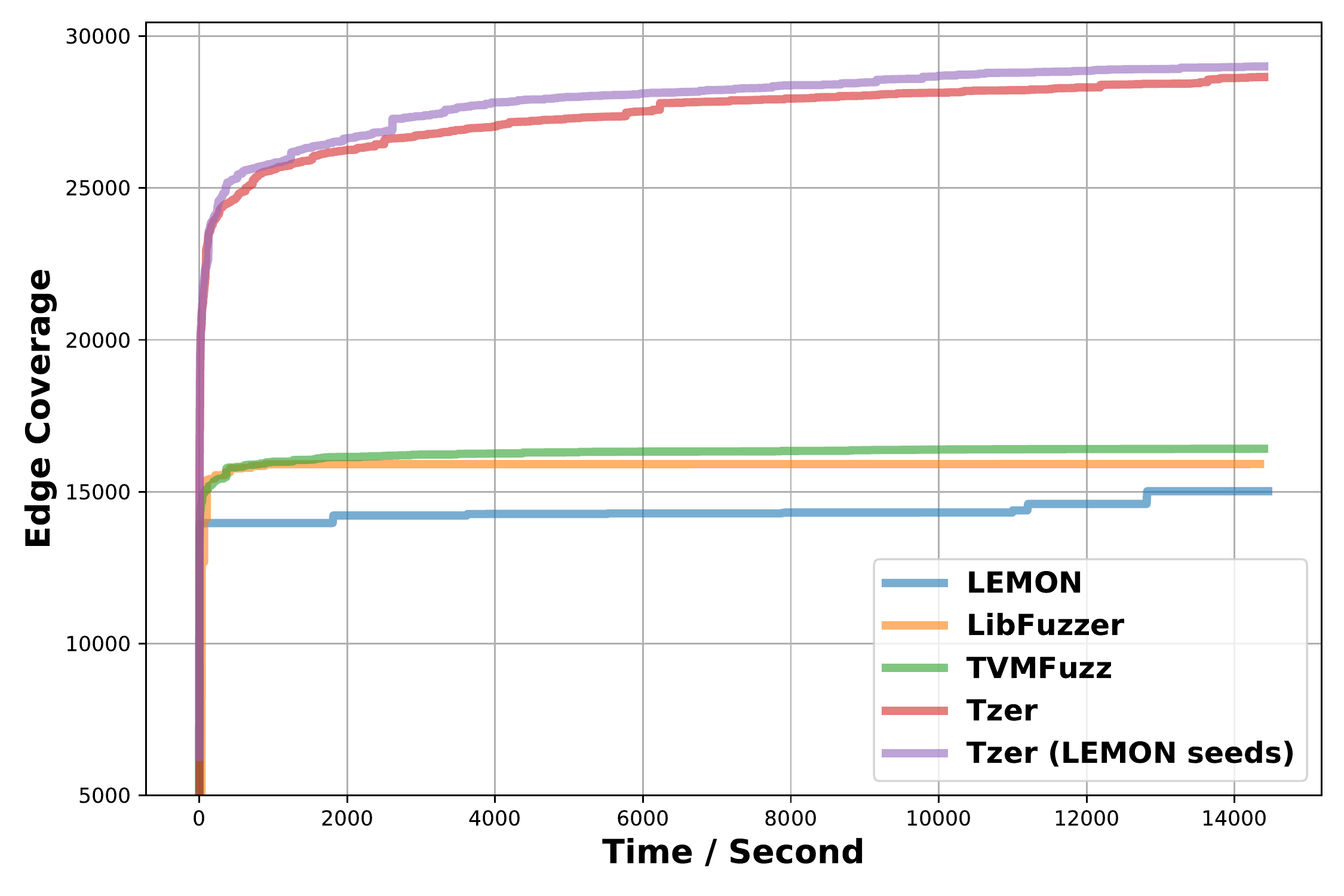}
    \caption{ Comparison with Existing Work}
    \label{fig:benmark_existing_wk}
\end{minipage}
\begin{minipage}{.48\textwidth}
    \includegraphics[width=\textwidth]{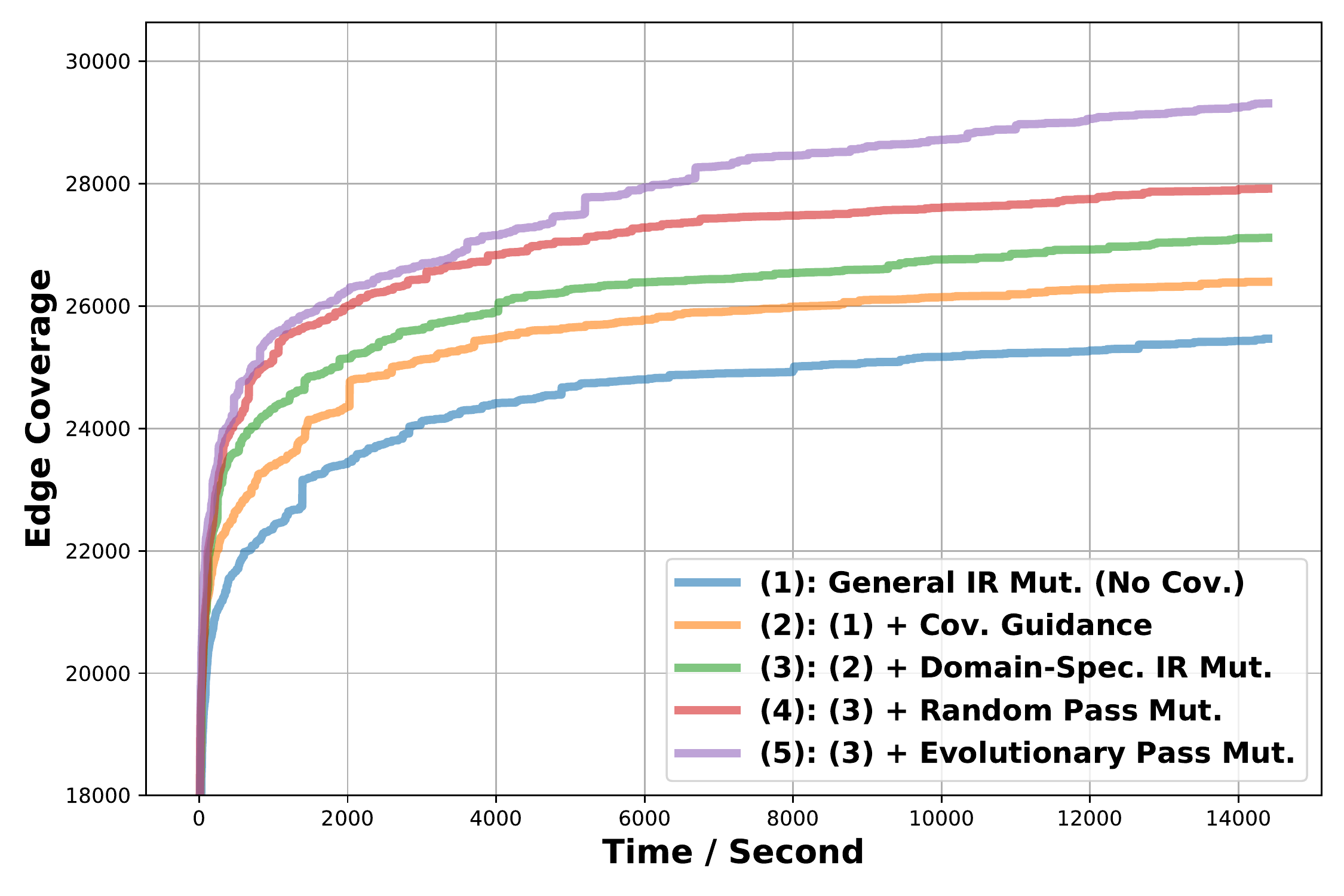}
    \caption{Ablation Study of \sys's Components}
    \label{fig:ablation_study}
\end{minipage}
\end{figure}

Figure~\ref{fig:benmark_existing_wk} presents the \emph{coverage trends} for both \sys and the compared existing work within the default 4-hour budget. 
To be specific, the $x$ axis presents the time costs and the $y$ axis shows the basic block coverage achieved.
More powerful techniques are expected to achieve higher coverage at the same timestamp.
As the figure shows,
\sys is able to beat other compared techniques at the very beginning and eventually achieves 75\% higher coverage than the 2nd-best baseline (i.e., \tvmfuzz).
Notably, \sys is able to keep visible coverage increase even at the late stage of the 4-hour budget while other techniques tend to converge very quickly. Another interesting observation is that \sys with the same seeds as \lemon even achieves slightly higher coverage than the default \sys, demonstrating the robustness of \sys.

Table~\ref{tab:valuable_tests} further presents the number of \emph{valuable tests} (i.e., the tests that are both compilable and able to trigger new coverage) generated by all the compared techniques within 4 hours.
Regarding the comparison of graph-level and low-level IR mutations,
\sys is able to generate 7.7x more valuable tests than the state-of-the-art graph-level mutator \lemon.
Specifically, \lemon only generates 63 valuable tests when the models are lowered to TIR functions (one model can be lowered to multiple TIR functions); if we had considered valuable tests at its original model level, the number of valuable tests is merely 20 out of all the 2.6k models generated by \lemon (i.e., 0.7\%).
We can also observe that LibFuzzer can hardly generate valid tests since it is a bit-level fuzzer, not aware of the grammar and semantics behind.
Lastly, among the low-level IR fuzzers, \sys is still able to outperform \tvmfuzz by 50\% in terms of valuable tests. The main reason is that \tvmfuzz follows a pure generation-based approach (which lacks coverage guidance and makes it challenging to simulate realistic IRs) and does not consider the mutual effect of IR and pass combinations.

\begin{table}
    \centering
    \begin{tabular}{c c}
\hline
         Tools & \# Valuable Tests \\
\hline 
\hline
         \sys                  & {\bf 497} \\
         \tvmfuzz              & 331 \\
         LibFuzzer             & 38  \\
         \hline
         \sys (\lemon seeds)   & {\bf 485} \\
         \lemon (\lemon seeds) & 63  \\
 \hline
    \end{tabular}
    \caption{Number of Generated Valuable Tests\protect\footnotemark}
    \label{tab:valuable_tests}
\end{table}
\footnotetext{If not specified, \sys seeds (\S~\ref{sec:exp:procedure}) are used by default. Initial seeds are not taken into account for fair comparison.}

\subsection{RQ2: Ablation Study of \sys}\label{sec:exp:rq2}

In this RQ, we further study the effectiveness of \sys's individual components:

\begin{enumerate}
    \item \textbf{RQ2.1}: Is coverage feedback helpful for tensor compiler fuzzing?
    \item \textbf{RQ2.2}: Can \pspec mutations further improve tensor compiler fuzzing?
    \item \textbf{RQ2.3}: Are pass mutations necessary for tensor compiler fuzzing?
    \item \textbf{RQ2.4}: Can our evolutionary joint IR-pass mutation (described in \S~\ref{sec:fuzzloop}) outperform a baseline joint IR-pass mutation that mutates both IR and pass sequences simultaneously? 
\end{enumerate}

To answer the above questions, we first build a simplistic variant of \sys that only applies general-purpose mutation (i.e., without coverage feedback, \pspec mutation, or joint IR-pass mutation). Then, we incrementally add more components to the simplistic variant in the order of coverage feedback, \pspec mutation, random joint IR-pass mutation, and evolutionary joint IR-pass mutation. 
Curves (1) to (5) in Figure~\ref{fig:ablation_study} represent the coverage trends after adding each component progressively.
From curves (1) and (2), we can see that coverage feedback has positive effects on tensor compiler fuzzing.
Curves (2) and (3) confirm the effectiveness of \pspec IR mutation in addition to general-purpose IR mutation.
RQ2.3 can be answered by comparing curve (3) against curves (4) or (5), as extended pass sequence mutation could help trigger more interesting behaviors.
Lastly, comparing curves (4) and (5), it can be shown that our evolutionary joint IR-Pass mutation is superior to the random joint IR-pass mutation, which performs coverage-guided fuzzing on IR files and supplies a randomly mutated pass sequence to each generated IR file.
Hence, we can draw a conclusion that all the main components of \sys contribute to tensor compiler fuzzing.

\subsection{RQ3: Parameter Sensitivity}\label{sec:exp:rq3}

\begin{figure}[h]
    \centering
    \includegraphics[width=\textwidth]{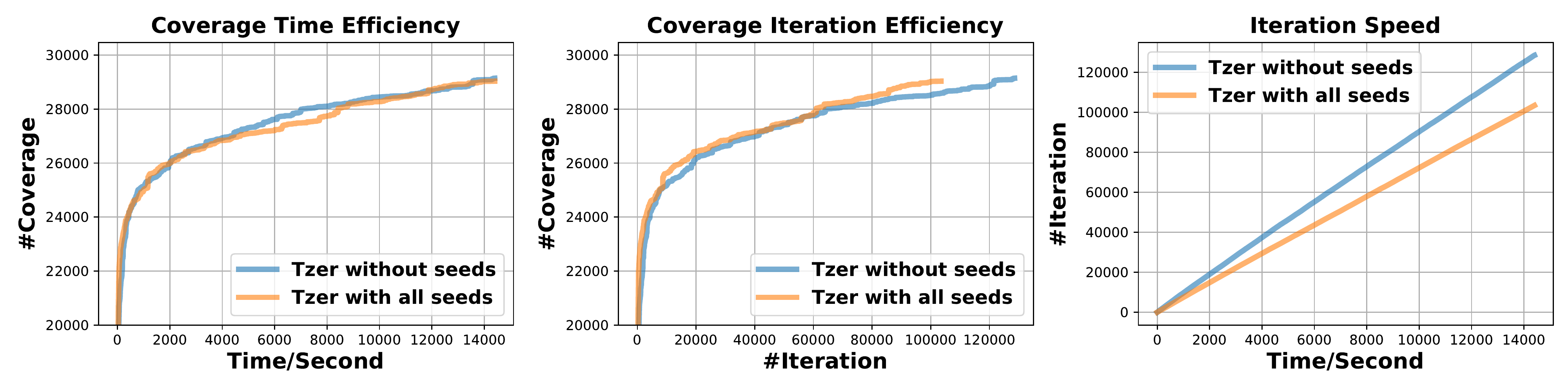}
    \caption{Impact of Seeds on \sys}
    \label{fig:exp:seeds}
\end{figure}

\noindent\textbf{Sensitivity to Seeds ($S_0$)}
The first sub-figure in Figure~\ref{fig:exp:seeds} shows how \sys performs with and without the default initial seed pool.
Surprisingly, the non-seed version has comparable (and even slightly better at some time stamps) effectiveness to the default \sys with 629 TIR seeds in terms of the coverage trend.
This is because, though each iteration \sys with seeds could generate higher-quality tests (the yellow curve is higher than the blue one in the 2nd sub-figure of Figure~\ref{fig:exp:seeds}), the non-seed version runs 24\% faster than that with seeds on average (as shown in the 3rd sub-figure).
The rationale behind is that if initial seeds are not given, \sys has to start IR mutation from an empty TIR function (i.e., \texttt{PrimFunc([]) \{0\}}) so that mutated variant IR files are similarly simple.
Hence, the overall compilation time of simple IRs will be smaller than the complex ones derived from real models.

\begin{figure}
        \centering
\begin{minipage}{.51\textwidth}
        \includegraphics[width=\textwidth]{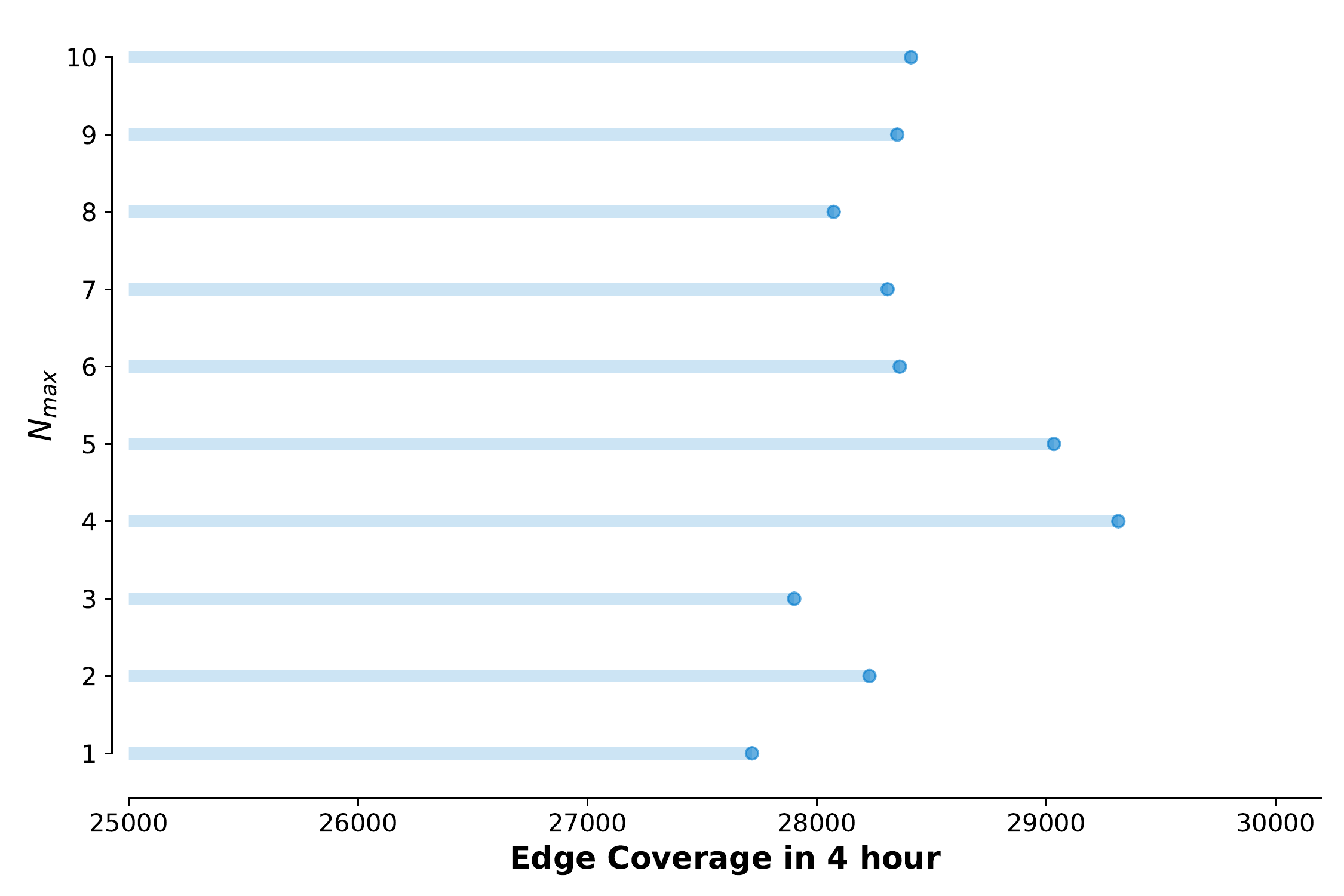}
        \captionof{figure}{Impact of Parameter $\NFail_{max}$ on Peak Coverage}
        \label{fig:rq3:nmax-peak}
\end{minipage}
\begin{minipage}{.47\textwidth}
        \includegraphics[width=\textwidth]{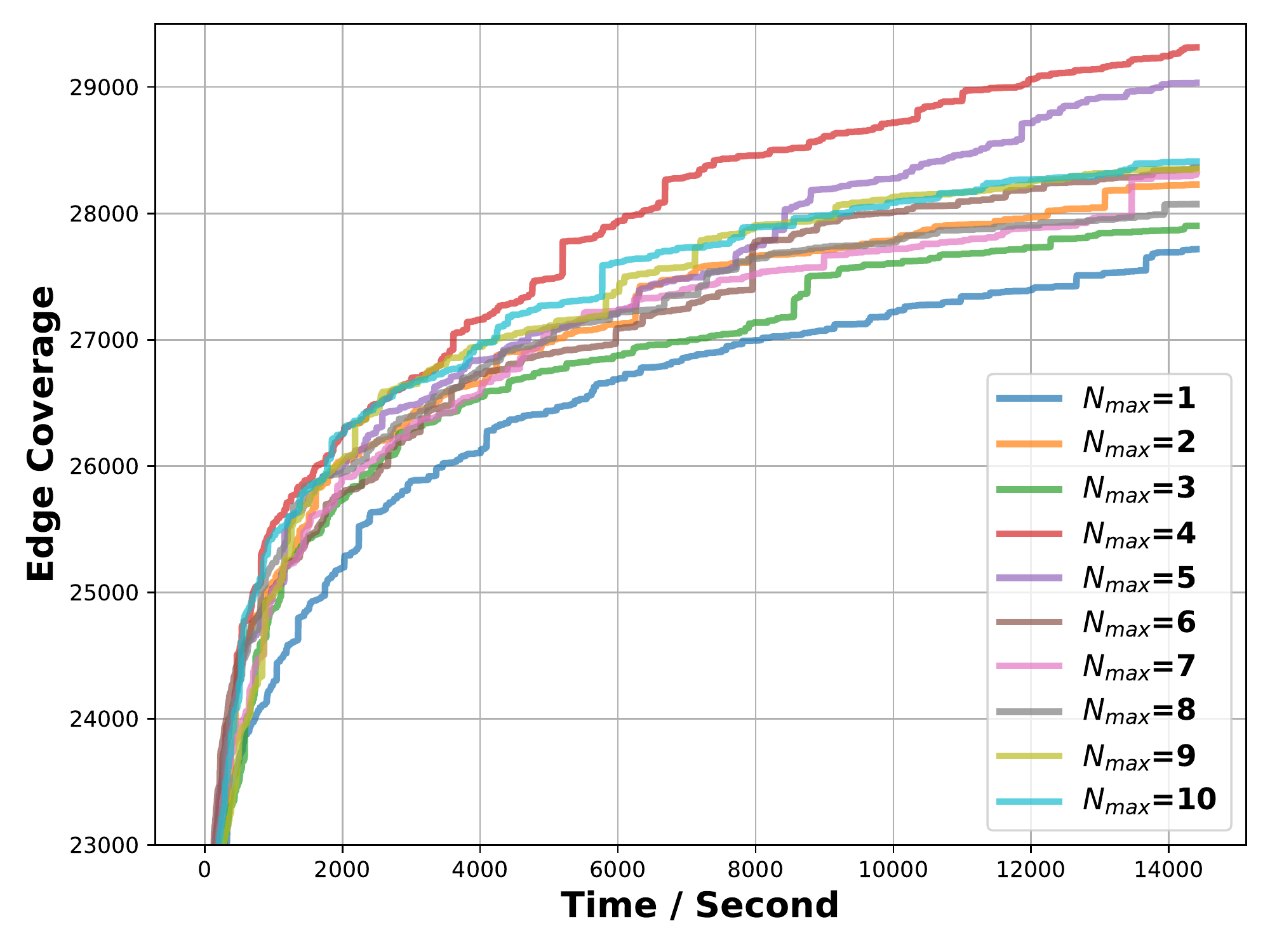}
        \captionof{figure}{Impact of Parameter $\NFail_{max}$ across Time}
        \label{fig:rq3:nmax}
\end{minipage}
\end{figure}

\noindent\textbf{Sensitivity to Pass Mutation Frequency ($\NFail_{max}$)}
Compared with traditional fuzzing loop, \sys has an extra parameter $\NFail_{max}$ to control the IR-Pass mutation interleaving (see \S~\ref{sec:fuzzloop}).
To study the impact of $\NFail_{max}$, we conducted the experiment using different values from 1 to 10 for $\NFail_{max}$.
Figure~\ref{fig:rq3:nmax-peak} presents the final 4-hour coverage of different settings, while Figure~\ref{fig:rq3:nmax} presents the corresponding detailed coverage trends. From the figures, we can see that $\NFail_{max} = 4$ demonstrates the best effectiveness.
In addition, $\NFail_{max} = 1$ performs the worse in terms of the peak coverage and overall trend.
This is because the coverage is mainly contributed by testing different IRs and the coverage growth will slow down if we frequently ``freeze'' the newly found IRs and mutate the pass sequences instead.
We can also observe that the coverage does not keep growing if we keep increasing $\NFail_{max}$ (i.e., decreasing the probability of pass mutation).
The rationale behind is that though pass mutation contributes less than IR mutation in the early stage, it is still important to mutate the pass sequence for an ``old'' IR that is not very likely to derive new interesting IRs anymore with its current pass sequence. \RevDel{In sum, the findings for this sub-RQ demonstrate the important trade-off for joint IR-pass mutation.}\RevAdd{}{In conclusion, it is important to distribute the frequency of pass mutation reasonably.}

\begin{figure}
    \centering
    \includegraphics[width=\textwidth]{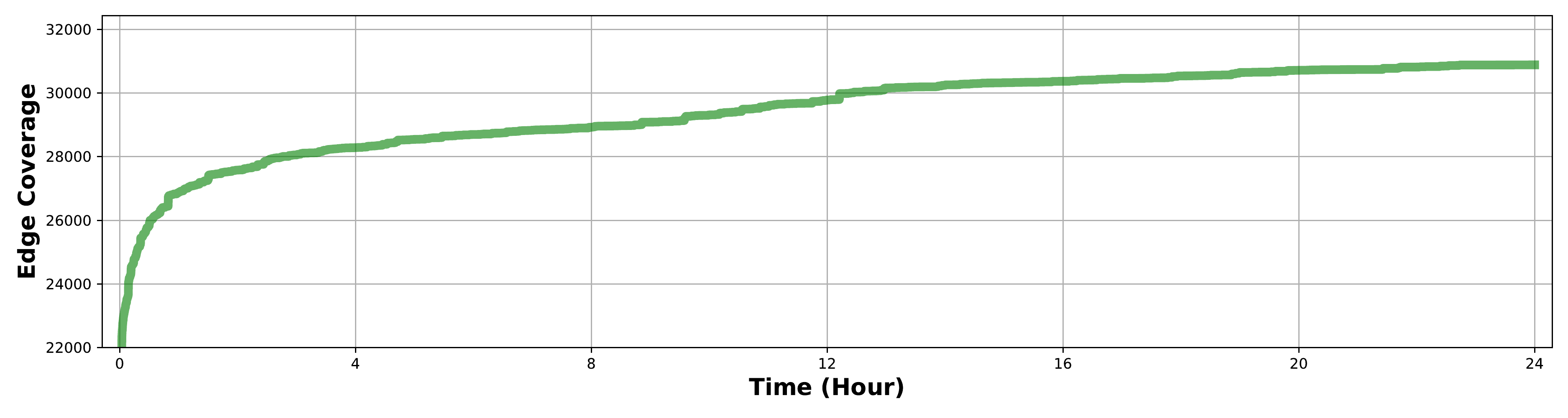}
    \caption{24-hour Coverage Trend of \sys}
    \label{fig:24hr}
\end{figure}

\noindent\textbf{Sensitivity to Fuzzing Time ($\Time$)}
Figure~\ref{fig:24hr} shows the overall coverage trend achieved by the default \sys across 24 hours.
While the existing techniques already saturate within 4 hours (shown in RQ1), \sys is able to successively keep coverage growth for the entire 24-hour period.
Specifically, the first 4-hour window contributes the most coverage, i.e., 91.6\%, while later 5 4-hour windows are still able to contribute 2.1\%, 2.8\%, 1.9\%, 1.1\%, and 0.5\% coverage respectively, demonstrating the effectiveness of \sys.

\RevAdd{(CHANGE LIST 5)}{In terms of the total code coverage of TVM, at the source code level, \sys can at best achieve 36.9\% line coverage and 28\% branch coverage with 4 CPU hours by only tracing the source files used in normal compilation. At the LLVM bitcode level, there are 482k CFG edges in total for our target, and \sys achieves about 6\% coverage within 4 CPU hours. This is because LLVM coverage sanitizer takes code bloating into account (C++ headers, templates and inlined functions are repeatedly considered) and thus can present underestimated coverage rates. Also, please note that 
modest overall coverage rates are quite common for fuzz testing of complicated software systems. For example, existing state-of-the-art Linux kernel fuzzers implement coverage collection with LLVM as well. Although they do not suffer from template code bloating in C++ as Linux is mostly implemented in C, the fuzzers can only achieve 0.8$\sim$10.5\% coverage after 50-hour fuzzing by fully utilizing a 32-core high-end CPU~\cite{kim2020hfl}.}

\subsection{RQ4: Bug Detection Effectiveness}\label{sec:exp:rq4}

To date, \sys has found \NBugs previously unknown unique bugs.
Table~\ref{tab:bug_sum} shows the detailed information about the \NConfirm bugs that have been confirmed by TVM's developers, where \NFix of them have already been fixed and merged to the main branch of TVM.

\sys generates tests through pass mutation, IR mutation, and their combination.
It is important to understand the necessity and effectiveness of each part. Table~\ref{tab:bug_compare} further presents the overall statistics for the bugs and bug types (categorized based on bug root causes) found by different studied techniques. \RevAdd{}{In terms of confirmed bugs}, we can find that in addition to only mutating IRs (i.e., Column ``\sys-IR''), 
modelling IR/Pass jointly (i.e., Column ``\sys-Full'') could help detect \RevDel{135.3\%}\RevAdd{}{2.17x} more bugs and \RevDel{66.7\%}\RevAdd{}{1.6x} more bug types.
Existing fuzzers for compilers, not limited to tensor compilers, only consider the compiler under test as a black box ingesting input source language texts and ignore the mutual effect of IRs and pass sequences internally applied together.
\sys demonstrates for the first time that it could be beneficial to perform evolutionary joint IR-Pass mutation for better and deeper bug detection.

From Table~\ref{tab:bug_compare}, we can also observe that bugs detected by \sys can hardly be detected by other compared techniques, e.g., \sys detects \MoreBug more \RevAdd{}{confirmed} bugs compared with the 2nd-best technique, \tvmfuzz.
This is mainly because \sys has a more complete modelling for both IR and pass sequences, as well as having a better fuzzing efficiency to quickly harness the large well-modelled search space (with coverage guidance).
In addition, according to Figure~\ref{fig:benmark_existing_wk}, \sys is able to consistently find uncovered CFG edge while other techniques converge at a very early stage, which explains why existing techniques fail to help discover more potential bugs.

\begin{table}[ht]
    \footnotesize %
    \centering
    \begin{tabular}{c|p{0.15\textwidth}|K{0.08\textwidth} K{0.08\textwidth} K{0.08\textwidth}|c|c}
        \hline \multicolumn{7}{c}{
\multirow{4}{*}{\parbox{\textwidth}{\centering
\textbf{API-I}: API Inconsistency (\S~\ref{sec:apii}).
\textbf{API-M}: API Misuse (\S~\ref{sec:apim}).
\textbf{AE}: Arithmetic Error (\S~\ref{sec:ae}).\\
\textbf{DL}: Driver Lifetime (\S~\ref{sec:dl}).
\textbf{FFI}: Foreign Function interface (\S~\ref{sec:ffi}). 
\textbf{IMA}: Invalid Memory Access (\S~\ref{sec:imc}).\\
\textbf{N/A}: (\S~\ref{sec:na}). 
\textbf{OOM}: Out Of Memory (\S~\ref{sec:oom}). 
\textbf{PMI}: Pass-Module Immutability (\S~\ref{sec:pmi}).
\textbf{TE}: Type Error (\S~\ref{sec:te}).
}}} \\
\multicolumn{7}{c}{}\\
\multicolumn{7}{c}{}\\
\multicolumn{7}{c}{}\\
        \hline
         & & \multicolumn{3}{c|}{Triggering Components} &  \\
        ID & Root Cause & IR & Pass & Runtime & Symptom & Status \\
        \hline \hline
        1 & TE      & \Y &    &    & Crash         & Fixed \\
        2 & IMA     & \Y &    &    & Crash         & Fixed \\
        3 & IMA     & \Y &    &    & Crash         & Fixed \\
        4 & TE      & \Y &    &    & Exception     & Fixed \\
5$\sim$7  & API-M   &    & \Y &    & Performance   & Fixed \\
        8  & API-I   &    & \Y &    & Exception     & Fixed \\ 
        9  & PMI     &    & \Y &    & Inconsistency & Confirmed \\ 
        10 & FFI     & \Y &    &    & Crash         & Fixed \\ 
        11 & API-I   &    & \Y & \Y & Crash         & Fixed \\ 
        12 & API-I   & \Y & \Y &    & Exception     & Fixed \\ 
        13 & IMA     & \Y &    &    & Crash         & Fixed \\ 
        14 & TE      & \Y &    &    & Crash         & Fixed \\ 
        15 & IMA     & \Y & \Y &    & Crash         & Fixed \\ 
        16 & IMA     & \Y & \Y &    & Crash         & Fixed \\ 
        17 & FFI     & \Y &    &    & Crash         & Fixed \\ 
        18 & FFI     & \Y &    &    & Crash         & Fixed \\ 
        19 & IMA     & \Y & \Y &    & Crash         & Fixed \\  
20$\sim$23 & AE      & \Y & \Y &    & Crash         & Fixed \\ 
24$\sim$26 & IMA     & \Y &    &    & Crash         & Confirmed \\
        27 & DL      &    &    & \Y & Crash         & Fixed \\
        28 & OOM     &    &    & \Y & Crash         & Fixed \\
        29 & N/A     &    &    & \Y & Exception     & Confirmed \\
        30 & IMA     &    &    & \Y & Crash         & Confirmed \\
\RevAdd{}{31$\sim$36} & \RevAdd{}{AE}      & \RevAdd{}{\Y} & \RevAdd{}{\Y} &    & \RevAdd{}{Crash}        & \RevAdd{}{Confirmed} \\ 
\RevAdd{}{37} & \RevAdd{}{AE}      & \RevAdd{}{\Y} & \RevAdd{}{\Y} &    & \RevAdd{}{Crash}         & \RevAdd{}{Fixed} \\
        \hline
    \end{tabular}
    \caption{Summary of Confirmed Bugs Detected by \sys in TVM (v0.8-dev). \RevAdd{}{Abbreviations like ``31$\sim$36'' represent multiple different bugs instead of one bug.}}
    \label{tab:bug_sum}
\end{table}

\begin{table}[h]
    \centering
    \begin{tabular}{c | c c c | c c }
         \hline
         Methods & \lemon & \tvmfuzz  & LibFuzzer & \sys-IR & \sys-Full  \\ 
         \hline
         \hline
         \#Valid bugs & 3 & 6 & 3 & \NIRBug & \RevDel{40}\RevAdd{}{37}\\
         \#Bug Type   & 3 & 5  & 3 & 6 & 10 \\
         \hline
    \end{tabular}
    \caption{Detectable \RevAdd{}{Confirmed} Bugs by Different Methods and \sys Components}
    \label{tab:bug_compare}
\end{table}

\subsection{Bug Root Causes and Case Study}

To demonstrate the versatility of \sys, we study all the 10 possible root causes of confirmed bugs detected by \sys as shown in Table~\ref{tab:bug_sum}, and discuss representative bugs for each category:

\subsubsection{Invalid Memory Access}\label{sec:imc}

Since most tensor compilers leverage memory-unsafe languages (i.e., C/C++) to implement the core components, it is not surprising that they will suffer from various memory problems, like out-of-bound access or \Null pointer dereference. 
Invalid memory access is one of the most frequently detected bug types by \sys.
For example,
\sys discovered an out-of-bound access bug triggered by a specific combination of IR and pass. 
When applying pass \texttt{InjectVirtualThread} on an IR module, 
the IR would be converted to the SSA format first, traversing all expressions to create a variable-to-array mapping for recording variable status. 
Theoretically, an array might not exist by variable name or it is temporarily empty.
However, \sys found during visiting \texttt{Save} or \texttt{Load} expressions, the TVM compiler only checks the existence of corresponding array and then directly accesses the last element (i.e., \texttt{std::vector::back} in C++) without boundary checking, resulting in a crash.
We illustrate a simplified fix to it in Listing~\ref{lst:oob}.

\begin{lstlisting}[language=C++, caption={Sample fix for Missing Boundary Checking\label{lst:oob}}, escapechar=@]
PrimExpr VisitExpr_(const LoadNode* op) final {
....
@\DelLine{\ \ \ \ if\ (scope\_.count(v))\ \{}@
@\AddLine{\ \ \ \ if\ (scope\_.count(v)\lfbox[hlgreen]{\ \&\&\ !scope\_[v].empty()}\})}@
         return Load(op->dtype, scope_[v].back(), op->index, op->predicate);
\end{lstlisting}

In addition to an out-of-bound access to containers, a crash occurs if a \Null pointer is dereferenced. 
According to the TVM design, objects could be nullable (an optional type containing a \Null state) or non-nullable.
In TVM, an object accesses its data members or member functions by the \texttt{->} operator in C++, which assumes any objects using operator \texttt{->} are not \Null objects.
However, even for nullable objects, \sys found over 44 functions (categorized as 3 unique bugs) do not check if a receiving nullable object is \Null or not, resulting in immediate crashes in case of \Null objects.

\subsubsection{Python-C++ FFI Handling}\label{sec:ffi}
Same as most other deep learning software, TVM and most other tensor compilers provide a Python interface, i.e., Foreign Function Interface (FFI), to bind Python functions and objects to C++ functions and objects through the \texttt{ctypes} standard library~\cite{ctypes} and \texttt{cython}~\cite{cython}. 
The motivation is that most deep learning practitioners are familiar with Python instead of C++.
However, python requires objects to support numerous built-in functions.
For example, \sys found the \texttt{StringImm} object in TVM failed to provide a \texttt{\_\_hash\_\_} implementation and threw an unexpected exception when put into a map container.

\subsubsection{Pass-Module Immutability} \label{sec:pmi}
TVM's passes mark the input IR module as \texttt{const} object (i.e., \texttt{const IRModule}), meaning that member functions that mutate data members cannot be called by such objects.
However, \sys found a pass, i.e., \texttt{ToBasicBlockNormalForm}, violating this contract by permitting the input \texttt{const} IR object to call non-\texttt{const} methods using pointers (C++ codebase), resulting in inconsistency issues in Python front-end.
We fixed this bug by forcing a copy at the beginning of the transformation.
A simplified bug fix is shown in Listing~\ref{lst:pass-mod}.

\begin{lstlisting}[language=C++, caption={Sample Fix for the Pass-Module Immutability Bug\label{lst:pass-mod}}, escapechar=@]
@\DelLine{IRModule ToBasicBlockNormalForm(const IRModule\& \lfbox[hlred]  {mod})   \{}@
@\AddLine{IRModule ToBasicBlockNormalForm(const IRModule\& \lfbox[hlgreen]{mod\_}) \{}@
@\AddLine{\ \ \ \ auto mod = IRModule(mod\_->functions, mod\_->type\_definitions, ...); // Deep copy.}@
\end{lstlisting}

\subsubsection{API Misuse}\label{sec:apim}
\sys also surprisingly detected that sometimes O4 optimization performs even worse than O2 (default optimization).
This is actually because we followed TVM's official tutorial when building \sys while their tutorial misused the API which failed to invoke the desired optimization.
In TVM's Python API, optimization level can be be identified within a scope called \texttt{PassContext} (Line 1 in Listing~\ref{lst:api-miss}).
In Listing~\ref{lst:api-miss}, old tutorial code calls \texttt{.evaluate()} outside the \texttt{PassContext} scope.
The \texttt{evaluate()} function, nevertheless, is where the optimizations are applied.
Therefore, when calling \texttt{evaluate} out of the O4 scope, the default optimization (O2) will be applied so that when comparing with another O2-optimized binary (they are all equally optimized), it is possible to see one is slower than the other due to uncertainty.

\begin{lstlisting}[language=python, caption={Sample Fix to API Misuse\label{lst:api-miss}}, escapechar=@]
with tvm.transform.PassContext(opt_level=4):
@\DelLine{\ \ \ \ executor = relay.build\_module.create\_executor("graph", mod, dev, target)}@
@\AddLine{\ \ \ \ executor = relay.build\_module.create\_executor("graph", mod, dev, target\lfbox[hlgreen]{, params).evaluate(})}@
....
@\DelLine{tvm\_out = executor\lfbox[hlred]{.evaluate()}(tvm.nd.array(data.astype(dtype))\lfbox[hlred]{, **params})}@
@\AddLine{tvm\_out = executor(tvm.nd.array(data.astype(dtype)))}@
\end{lstlisting}

\subsubsection{API Inconsistency}\label{sec:apii}
Inconsistency in API happens when a program does not act as what the API is specified.
For example, when running programs on heterogeneous devices (e.g., run a program that requires both GPU and CPU),
TVM splits the functions into either the host side or the device side.
There is a parameter controlling the calling convention (i.e., \texttt{calling\_conv}) for the heterogeneous compilation, which is set to \texttt{kDefault} by default.
\texttt{kDefault} generally means that both host and device targets are CPUs (e.g., LLVM as the target).
However, a pass called \texttt{DecorateDeviceScope} violates the calling convention by implicitly change \texttt{kDefault} into \texttt{kDeviceKernelLaunch} which is built for non-CPU device targets (i.e., the \texttt{DecorateDeviceScope} is not desired to change the calling convention).
Such an inconsistency leads to a crash at runtime.

\subsubsection{Type Error}\label{sec:te}
\sys found an issue regarding TVM's constant folding in integer conversion.
For example, the expression \texttt{assert tir.const(1) == tir.const(True)} would throw unexpected exceptions, whereas we expected it to be a \texttt{True} after evaluation.
The root cause is that conversion for signed/unsigned integers (int64 and boolean) is not well handled.
Theoretically, since the range of boolean type is the subset of int64's, we can convert the boolean value to an int64 value.
We fundamentally fixed the issue by refining TVM's type conversion for signed and unsigned integers.

\subsubsection{Driver Lifetime Error}\label{sec:dl}
\sys found that when enabling CuDNN~\cite{cudnn} as the target backend, TVM crashes after being stuck for a while when the program exits.
This is because TVM made the CuDNN device handler of a whole-process lifetime by marking it with \texttt{thread\_local} (a specifier in C++).
Thus, according to the RAII rule~\cite{raii} of C++,
the deconstructor to release the handler will be called during program exit.
However, the CuDNN library context might have already been exited when such release handlers are being called, causing segmentation fault after a long suspension.

We further proposed 2 fixes to this problem:
(1) we register the handler release function at exit time using \texttt{atexit}, and make sure that the destroyer of library context will be called after it; 
(2) we simply remove the handler release code and let it leak since we do not need to do recycling when a program is going to exit.
The community finally accepted proposal (2) since proposal (1) is more advanced and complex, increasing the maintenance cost.

\subsubsection{Out-of-Memory}\label{sec:oom}
\sys found an interesting out-of-memory (OOM) bug when using the virtual machine (VM) as TVM's executor.
The cause is that the previous VM memory allocator never releases occupied memory in the  memory pool and leverages no memory defragmentation strategies. 
It only re-uses memory blocks in the pool if the incoming request size is smaller than the existing one.
When the memory requests follow a monotonic pattern, it fails eventually since it cannot release previous memory blocks in the pool.

For example, as shown in Table~\ref{tab:oom_exp}, on a GPU of 8 GB memory, if for each time, we release $i$ GB memory and allocate $i+1$ GB memory ($i$ starts from 0), it will fail in the 4th step.
The reason is that each time, after releasing $i$ GB memory, the released memory chunk is returned to the free list; when requesting $i+1$ GB next time, all chunks in the pool cannot be used since they are smaller than $i+1$ GB. 
Hence, in the 4th step, even though the GPU has 8 GB physical memory, it cannot allocate a 4 GB memory chunk.

We fixed this issue by simply releasing all cache blocks and re-attempting allocation if any OOM exceptions are caught.

\begin{table}
    \centering
    \begin{tabular}{c|c | c c c}
         \hline
         \#Iter. & Next Action                   & Avai. System Mem. & Occupied Mem. & Free list (pool) \\
         \hline
         \hline
         1       &  Alloc. 1 GB                  &   8 GB            &       0       & empty       \\
         2       &  Free 1 \& alloc. 2 GB        &   7 GB            &       1 GB    & empty       \\
         3       &  Free 2 \& alloc. 3 GB        &   5 GB            &       2 GB    & [1] GB      \\
         4       &  Free 3 \& alloc. 4 GB (Fail) &   2 GB            &       3 GB    & [1, 2] GB \\
         \hline
    \end{tabular}
    \caption{Example of How TVM's VM Allocator Failed in Monotonic Allocation.}
    \label{tab:oom_exp}
\end{table}

\subsubsection{Arithmetic Error}\label{sec:ae}
\sys also found some functions in TVM fail to check the legality of arithmetic operations, such as division by zero. This bug lies in an optimization that simplifies the calculation of TIR. Specifically, when TVM tries to simplify a division expression whose two operands are of type Ramp and Broadcast, it will directly modulo two numbers without checking the divisor. This causes the program to crash when the divisor is 0.

\subsubsection{Unknown Failure}\label{sec:na}
Among the bugs found by \sys, there is one whose root cause is still unknown to date.
The phenomenon of the bug is exception fleeing.
For example, when the device is running out of memory, it is expected to throw an OOM exception throughout the call stack to form a failure trace logging.
However, we found that sometimes, though thrown, the exception disappears during stack unwinding and allows the error program to continue the execution.
Since the continued program is incorrect, it eventually fails elsewhere with another exception  instead of the root exception.
When debugging it with GNU Debugger~\cite{gatliff1999embedding} or LLVM Debugger~\cite{lee2013using} to monitor the exception path, it strangely skipped \texttt{\_\_cxa\_catch}~\cite{cxaabi} and just fled away.
Though this strange bug can be reproduced on different operating systems and compilers and has been confirmed by TVM's developers, it still remains unfixed at this moment.

\section{Conclusion}

The evolution of tensor compilers requires automated testing to achieve high maintainability and reliability.
We demonstrate that existing fuzzing techniques are not tailored or effective enough to fulfill this mission.
To this end, we present \sys, a practical coverage-guided tensor compiler fuzzer with joint IR-Pass mutation.
Unlike traditional compiler fuzzers, \sys performs joint IR and pass mutation to explore various program states and introduces coverage guidance to navigate the mutation process.
Specifically, in addition to general-purpose mutators, \sys also leverages tailored \pspec mutators to target the hotspot logics behind tensor compilers.
The evaluation shows that \sys substantially outperforms the state-of-the-art fuzzers including a general-purpose fuzzer (i.e., LibFuzzer), a graph-level DL model fuzzer (i.e., \lemon), and the only domain-specific fuzzer for TVM (i.e., \tvmfuzz).
As one of the practical contributions of \sys, to date, we have helped the TVM community find \NBugs new unique bugs, with \NConfirm confirmed and \NFix of them already fixed in the current TVM version.
Our effort has been highly recognized by the TVM community, and the leading author of \sys has been nominated as a community reviewer for TVM.

\bibliographystyle{ACM-Reference-Format}
\citestyle{acmauthoryear}
\bibliography{reference}

\end{document}